# Some Properties of the Random Universe[a]

Anthony E. Scoville
Aerovest, Inc.
36 Taconic Road, PO Box 373, Salisbury, CT 06068
e-mail: anthonyscoville@aya.yale.edu

## Abstract

*What is the role of the constants of nature in physical theory? I hypothesize that the observable universe, $u_0$, constitutes a Universal Turing Machine (UTM) constrained by algorithmically random logical tape parameters defining its material properties (a physical UTM). The finite non-zero empirical values of Planck's constant, $h$, and other constants of nature exemplify those logical parameters. Their algorithmic randomness is necessary and sufficient for the consistent operation of a physical UTM. At any given time, $t_i$, these constants correspond to the first $n$ random halt digits, $\Omega_n$, of Chaitin's Halting Probability, $\Omega$. Planck's equation $E = h\nu$ and Boltzmann's relation $S = k_B \log_2 W$ are shown to apply to the operation of a physical UTM. The genomic evolution of $u_0$ in constants of nature space (CON space) from an undecidable state in $u_0$'s Planck era to its current ordered condition occurs through the algorithmically random, symmetry-breaking addition of new constants to the laws by which $u_0$ operates — a process called logical tunneling. The temperature of $u_0$ for $t \leq t_p$ is shown to be $T = 0°K$ The energy dissipated when a physical UTM clears its memory after each computation is proposed as a candidate for cold dark matter (CDM) and is calculated to comprise 87.5% of the matter content, $\Omega_M$, of $u_0$. This result concurs with current astronomical estimates that 87% of the matter content of $u_0$ consists of CDM. The energy incorporated in $u_0$ through the process of logical tunneling from undecidable states of the complete Universe, E, to decidable states of $u_0$ is suggested as a candidate for the unexplained "dark energy", $\Omega_X$, hypothesized to drive the accelerating cosmological expansion of space and believed to constitute 66% of the critical mass, $\Omega_0$, of the observable universe.*

---

# Contents

**I. Unsolved Problems in Physics:** *Why is physical reality quantized? Why is mathematics the language of physics? Can laws of physics explain the constants of nature?*

Quantum mechanics and its extension, quantum field theory, are powerful theories but no one knows why they work. Planck's quantum hypothesis and the quantization of quark-gluon fields by his successors are inspired guess-work[1] but efforts to quantize gravity have been unsuccessful. While string/M-theory may succeed in quantizing General Relativity, even if it demonstrates its mettle as "Theory of Everything" (TOE), the theory provides no explanation as to why our universe is characterized by a fundamental length scale of some particular value.[2]

Is there some property of *any* mathematical theory of physics that requires the quantization of physical reality? Why do Planck's quantum and other constants of nature have their observed values? Why is mathematical reasoning such a powerful tool for expressing the systematic relationships among phenomena studied by physics and other natural sciences? The evidence is overwhelming: nature appears to obey mathematical laws *everywhere* and for *all* time. So far as is known, constants of nature, including Planck's constant, are inexplicable "facts" and mathematics simply works extraordinarily well in the physical sciences.[3,4,5] Is the "unreasonable effectiveness of mathematics in the natural sciences"[6] evidence for a principle of physics in its own right? Is there a "reason" why the constants of nature are inexplicable facts?

Herein I argue that the foregoing problems, the first concerning the quantization of physical reality and the values of the constants of nature and the second concerning the efficacy of mathematics in the natural sciences, are siamese twins: quantization and algorithmically random constants of nature are necessary and sufficient for the laws of physics to operate consistently and therefore to be mathematically expressible. Some cosmological consequences of randomness in the laws of physics are also explored.

**II. Proposal:** *Turing Mechanics*: *Any Physical System is a Mathematical Machine*

To account for the striking correspondence between mathematical thought and our experience of physical reality, I propose that the universe itself constitutes a mathematical theorem-proving machine — a computer. Is that hypothesis supported by observational evidence? The hypothesis extends the formal theory of the Universal Turing Machine and its mathematical equivalents, Alonzo Church's λ calculus and the theory of general recursive functions, to physical systems defined by arbitrary, that is, *algorithmically random*, parameters specifying the material properties of any particular computer.[7,8]

> **Hypothesis: Principle of Mathematical Physics** (PMP): *Every law of physics and every state of a physical system corresponds to some program of a Universal Turing Machine (UTM).*

The Church-Turing thesis asserts that every computer is equivalent to some program of a Universal Turing Machine.[9] PMP is the Church-Turing thesis applied to the physical world. Therefore:

> **Theorem II-1**: *The rational and observable universe, $u_0$, and every subsystem thereof is a computer program that can be simulated by a Universal Turing Machine.*

The physics derived from PMP and Theorem II-1 will be called *Turing Mechanics* (TM).

The Principle of Mathematical Physics is motivated by four considerations. First, it accounts for the uncanny efficacy of mathematics in modeling the physical world. Second, in all of mathematics and physics there is only one "theory of everything" that has survived every challenge to date: namely, Turing's theory of computation and its formal counterparts embodied in the works of Church and Gödel[10], as extended by Gregory Chaitin[11,12,13,14,15]. In the attempt to construct a "theory of everything," it is seldom considered what one means by everything and whether that "everything" is the same as "rational" or computable. Turing's theory of computation is the prime mathematical candidate for a physical theory of everything not only because the theory is demonstrably universal but also because Turing's and Gödel's work defines "everything" precisely in terms of completeness. Turing's theory specifies the properties that a physical theory of everything must possess if it is to model the relationship between completeness and rationality in the universe. It is, therefore, rich enough to permit an isomorphism between our mental conception of the universe and the physical universe itself.[16] Third, the mathematics of all currently accepted theories of physics breaks down in some domain — notoriously in the Planck domain believed to characterize the nascent universe. The works of Turing, Gödel and Chaitin specifically address the issue of what happens when a mathematical formalism reaches its limits as one crosses the boundary between *decidable* theorems and *undecidable* theorems that can be neither proved nor disproved within mathematics. Thus their work may enable us to understand what happens logically when equations of physics yield a singularity. Fourth, because their work establishes the existence of true but undecidable theorems in mathematics, one can model the logical relationship between arbitrary constants of nature and the dynamic transformations of physical reality that we symbolize in the laws of physics specific to our universe, $u_0$.

The existence of undecidable theorems in mathematics requires one to distinguish between the *observable or "rational" universe*, $u_0$, consisting of the evolution of all physical systems whose states can be computed by an algorithm consisting of a finite number mathematical relationships expressed in a finite number of symbols and the *complete Universe*, $E$, consisting of $u_0$ plus all states that are consistent with $u_0$ but which can never be computed by its algorithms. Given the incompleteness of mathematics, PMP implies that the rational universe modeled by finite laws of physics is only part, indeed a very small part, of the complete Universe. (Figure II-1)

**Figure II-1**

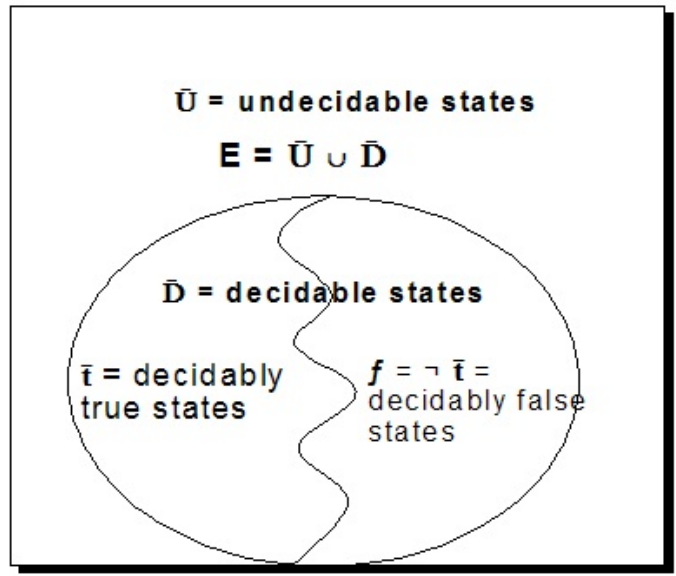

The observable universe, $u_0$, comprises all decidably true states $\bar{t}$ while the decidable universe comprises $\bar{t}$ plus all decidably false states $f$. As will become apparent from Figure IV-1 where I model the evolution of $u_0$ in *constants of nature space* (CON space), relative to $u_0$, $f$ relative to $u_0$ consists of all points not on the trajectory of $u_0$ in CON space. Throughout this paper *"universe" (lower case) always refers to $u_0$; "Universe" (capitalized) always refers to E.*

The contrast between $u_o$ and $E$ can be stated precisely in terms of the following definitions:

> **Definition II-1**: *The computable or decidable ("rational") formulae of a system is the set of formulae for which there exists an algorithm (law of physics) capable of formally generating any member of the set or its negation by the application of the algorithm in a finite number of steps.*

Notice that Definition II-1 requires that a UTM always halt on some member of the computable set or its negation.

**Definition II-2**: *The recursively enumerable (listable) formulae of a system is the set of formulae for which there exists a finite algorithm (law of physics) capable of formally generating any member of the set by the application of the algorithm in a finite number of steps.*

Recursive enumerability is a weaker condition than computability in that it only asserts which formulae are members of the set; it asserts nothing about determining which formulae are not computable by the algorithm.

**Definition II-3**: *A formula is said to be formally undecidable if neither the formula nor its negation can be formally generated by the laws of physics in operation at any given time in a finite number of steps unless that formula is assumed as an ad hoc axiom of those laws.*

**Definition II-4**: *A set of formulae is said to be algorithmically random if each member of the set is undecidable relative to any other subset of the formulae.*

As will be discussed in §V, a sequence of formulae is also algorithmically random if the number of bits addressing the shortest program capable of generating all members of the set is equal to the number of bits required to address all members of the sequence.

In terms of Definitions II-1, through II-4 the states of $u_o$ are computable by the algorithms defining $u_o$; the states of $E$ consist of all the decidable states of $u_o$, $\bar{D}$, plus all states, $\bar{U}$, that are formally undecidable by the algorithm(s) defining $u_o$.

It is often asserted that quantum computers can perform computations that are impossible to carry out on a Universal Turing Machine.[17] Such is not the case. A quantum computer can be modeled as the Feynman path integration of an ensemble of Turing machines operating in parallel along each of the virtual particle paths. It can be shown that any set of parallel Turing machines is equivalent to a single UTM.[18] I am not saying that a quantum computer cannot perform certain computations faster than computers constructed as a classical Turing machine. Rather I am only asserting that when a UTM is supplied with the material principles of a quantum computer, then it can perform any computation that can be performed by a quantum computer.

Any particular computer is specified by material parameters. These parameters distinguish, for example, an abacus from an electronic computer. Similarly, any law of physics requires the specification of parameters appropriate to the type of system being considered. These parameters must be incorporated in the program of a Universal Turing Machine in order to perform any computation. When the parameters are included as input, a UTM can simulate the behavior of a physical system operating under *any* set of dynamical laws[19,20] — so long as one subscribes to PMP which requires that all laws of physics be mathematical in the sense defined by Church and Turing.[21,22]

A UTM is a program; it is software, not hardware.[23] Turing machines define formal relationships between symbols that are themselves meaningless; they are purely syntactical programs. A UTM can include lines in its program specifying the physical parameters characterizing the hardware upon which a program is to operate. Indeed, the program must specify these parameters; otherwise one cannot perform any particular computation. Even though these hardware parameters must be included in the program, they do not determine what computations a UTM can perform. An abacus, an electronic computer and a quantum computer can each compute any natural number, provided that the number is computable. Thus, each can operate as a Universal Turing Machine; however, specific values of the hardware parameters certainly affect how fast one can solve any particular problem. Definition II-5 reflects this situation:

> **Definition II-5**: *A Theorem Proving Machine* (*TPM*) *is a Universal Turing Machine whose program includes values of the hardware parameters defining the particular physical machine upon which a computation is performed*.

In §III I briefly describe a Turing machine and suggest that the hardware parameters referred to in Definition II-5 characterize the properties of the tape upon which a UTM operates when reading and writing a program. In §IV I shall argue that some possible values of the hardware parameters are not permissible on logical grounds. In §V I shall argue that the specific permitted values of these parameters can never be computed by formalized Peano arithmetic; they must be added as independent axioms to the Peano axioms.[24]

Employing Definition II-5, Theorem II-1 can be stated more precisely:

**Theorem II-2**: *The universe, $u_0$, and every subsystem thereof is a theorem proving machine, TPM*.

Theorems correspond to states of a system or of a subsystem thereof; the "proof" of a state corresponds to the transformation of some given "initial" state (axiom or theorem as the case may be) into the designated state by the application of a precisely defined rule (recursive function) to the initial state. When a TPM proves a state $s_i$ from some other state $s_{i-1}$, the program $P: s_{i-1} \rightarrow s_i$ is said to "halt" on the state $s_i$.

## III. The Turing Machine

The following remarks concerning Turing machines borrow closely from John Casti's presentation.[25]

Turing machines constitute the most general of the four types of language acceptor in the Chomsky hierarchy.[26] All computers have two components: a tape upon which sequences of symbols are written and a black-box language acceptor (LA) that transforms symbols on the tape (input data) into new symbols on the tape according to precise rules. (Figure III-1.)

**Figure III-1**

Turing Machine

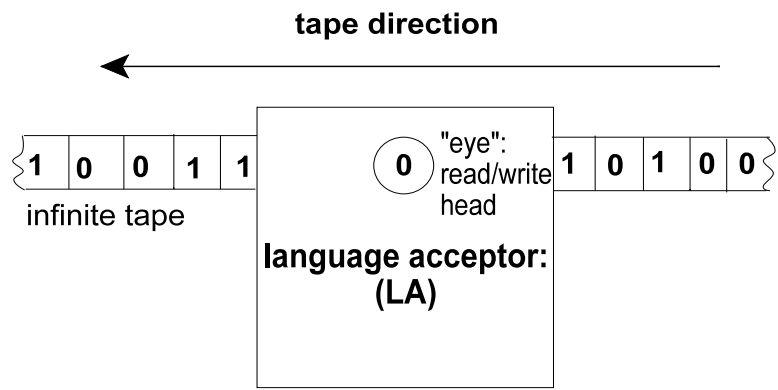

The transformation of a symbol on a tape requires three steps: first LA accepts or "reads" the symbol on the tape (in this instance the "**0**" under the "eye"); second, LA erases the existing symbol on the tape momentarily creating a blank space on the tape; third, LA writes some new symbol on the (momentarily) blank space and moves the tape one space to the left or right depending on the machine's program.[27] The program may also contain branching instructions that cause LA to move one or more tape spaces to the right or left depending what symbol is being read by LA. A branching instruction does not alter the symbol under the read/write head. Upon reaching some desired tape, state LA halts operation and the theorem corresponding to the state of the machine is said to be "proved." Any sequence of such transformations, including the instruction to "halt," is called a *Turing-Post Program*.[28] Branching statements can be eliminated by allowing LA to possess a finite number of *internal* states $\{A_i\}$. A program can then be written as a table describing the action to be taken for each possible symbol-state combination.

Any given table corresponds to just one program among the countable infinity that might be constructed to perform specific tasks. Turing's insight was to construct a *standard* program that could interpret (read and translate) a suitably coded version of *any* particular program *supplied as input data on its tape.* The *standard* program can then simulate the actions of any other program. The standard program is called a *Universal Turing Machine*.

From Theorem II-2 it follows that formal systems, computer programs and dynamical systems in physics are equivalent. The following table (after Casti)[29] summarizes the relationships among them:

Table III-1

| Formal System | Computer Program | Dynamical System |
|---|---|---|
| abstract symbols | tape symbols (marks) | number field |
| possible symbol strings | possible tape patterns | state manifold (phase space) |
| symbol string | tape pattern | state (point in phase space) |
| grammar (formation rules) | set of admissible tape patterns | constraints (prohibited regions in phase space) |
| starting axiom(s) | input tape pattern | initial state |
| logical rules of inference | Turing-Post programs | vector (force) field |
| proof sequence | sequence of tape patterns | trajectory (curve) in phase space |
| theorem | final tape pattern when the program halts or goes into an infinite loop | attractor (relative to prior points in trajectory) |
| symbol and string separation rules | tape parameters: size of tape segments, tape speed, tape strength | symmetry-breaking parameters: constants of nature |

The first eight rows of Table III-1 characterize the universal properties of formal systems, computer programs and dynamical systems which enable these systems to compute any computable number. The bottom row designates those hardware parameters that define any particular system when the universal properties of all such systems are realized in a specific (physical) computer or dynamical system. For the remainder of this paper I shall consider whether any logical requirements limit the tape parameters to certain permissible values.

**IV. Consistency, Objective Laws of Physics and Quantization**

The most important property characterizing any physical system is that it operate consistently. If the laws of nature harbor an inconsistency, then *every* formula would be a theorem formally provable from those laws.[30] Were such the case, it would be impossible for us to test (falsify) a theory of physics by experiment *or any other means*. What we might believe to be objective knowledge would constitute only personal prejudice. More generally, no TPM and no physical system could decide, by any conceivable rules of inference or process of computation, what its next state should become given its current state. No objective laws of nature could exist. In this sense, all properties of a physical system derive from the consistency of the laws governing its behavior.

For example, if Planck's radiation law were based on an inconsistent set of axioms, you could assert that the temperature of the photosphere of the sun is $5,800°K$; I could assert that it is $10^6 °K$ or even $0°K$. Both of us could derive our conflicting conclusions from Planck's law and it would be impossible to design an observing instrument to distinguish between these conflicting assertions.

In mathematics one can require, by methodological fiat, that any argument leading to an inconsistency is to be rejected for the reason that concept of proof breaks down. One is free to mandate this procedure because mathematics constitutes a human artifact that we construct according to our purposes. If the rules of procedure are accepted by all players in the game of mathematics, that is what mathematics is. In physics, one does not have this liberty and neither does any physical system — per Theorem II-2. If we are to employ mathematical reasoning to understand the physical world, one must find empirical evidence for the consistency of nature.

Observed phenomena that are consequences of suspected laws of physics are not acceptable as evidence for the consistency of these laws. Employing the regularities of nature to confirm the latter's consistency assumes the very consistency that one seeks to demonstrate. Rather, one requires evidence for some property of the physical world that makes consistency possible.

The ability to prove every formula from an inconsistency has a converse with profound implications. If the laws of physics operate consistently, there must exist at least one state of nature (formula) that is not derivable from those laws. Therefore one must demonstrate at least one fact (formula) of nature that one can prove to be *undecidable* from *any* conceivable laws of physics. From Theorem II-2 it follows that any qualifying demonstration depends upon the universal characteristics of effective computation embodied in the operation of a TPM.

The demonstration will proceed in three stages:

- In this section it will be argued that quantization of the Turing machine tape is necessary for the consistent operation of any TPM.

- In §V it will be shown that the particular values of the tape quantization parameters in any universe are algorithmically random. It follows that their values, *if they can be determined*, are sufficient to establish the consistent operation of any mathematical laws of physics.

- In §VII I shall suggest that the empirical constants of nature $h$, $G_N$, and $c_{vac}$ correspond to the tape parameters characterizing the TPM, $u_0$, at the Planck scale. Experimental determination of their finite, non-zero values demonstrates that the model of $u_0$ as a TPM is not vacuous and constitutes evidence necessary for the consistent operation of all laws of physics for $u_0$.

To begin, consider the Turing machine tape. (Figure III-1) Overprinting of one symbol on another is not permitted. It follows that the tape must possess three material properties. The first is *quantization*. The tape must be divisible into segments of finite non-zero *length*, *h*, in some appropriate phase space so that symbols can be unambiguously printed on it and can be completely erased from it.  The length must be greater than zero; otherwise, symbols printed by TPM would spill over from one segment to another and overprinting would result.  Conversely, unless the symbols themselves were infinitesimal, LA would only read an infinitesimal fragment of each symbol on each zero-length segment as input to its operations and could not interpret its input.  If, however, the symbols were infinitesimal, then the quantity of printer's ink required to emboss one symbol on the tape would differ by an arbitrarily small amount from the amount of ink required to emboss some other symbol.  Thus any given symbol would be indistinguishable from its adjacent symbol.[31]  Similarly, the length of a tape space cannot be infinite.  If it were, no matter how far LA moved under its reading head, LA could never scan all the printer's ink embossing the symbol and LA could not be certain that the symbol it unambiguously distinguished the symbol it is reading from any other symbol on the tape — which is equivalent to requiring the quantity of printer's ink embossing each symbol to differ from symbol to symbol by an arbitrarily large amount.

The second property is a *maximum tape velocity*.  The tape must move at most a finite number of segments for each computation completed by TPM.[32]  The tape cannot shift an infinite number of segments between computations because the presence of infinitely distant branching instructions is equivalent to providing LA with an infinite number of internal states.  In that case, internal states would correspond to an alphabet of indistinguishable primitive symbols differing by an arbitrarily small extent from each other.[33]  This situation implies that, when the tape happens to advance in one direction only, there is a maximum finite velocity, *c*, at which *changes* in the computational results printed on a segment can propagate along the tape: namely, a shift of one tape segment per computation.

The third tape property is *logical cohesion*.  The tape must not "tear" under *any* operating conditions to which it is subjected when moved through the language acceptor.  Were it to do so, a TPM's program could split into subroutines of arbitrary length at least one of which might lack the essential instructions, *U*, required to carry out any computation. In that situation, there would exist at least one (fractured) algorithm incapable of distinguishing between one symbol and another.

Of the three tape properties, logical cohesion, unlike quantization and maximum velocity, seems "unphysical" in that it constitutes a formal property of arithmetic. The dichotomy is only apparent. Given PMP every formal property corresponds to physical property and vice versa. The formal property expresses the requirement that, if a theorem is proved once, then it must be possible to reprove the theorem at any later time. Logical cohesion, manifest as a constant *G*, expresses the coupling between the axioms and the theorems of any formal system. Physically, a Turing machine tape can never tear and every state of a system is irrevocably coupled to every past state of the system.[34]

In sum, the distance by which TPM advances its tape between computations must be finite and non-zero corresponding to the physical manifestation of one bit of information. Therefore, if one defines $\ell$ to be the length of a tape segment, then $\{\ell \in \mathbb{N} \mid 0 < \ell < \infty\}$. Rational values of $\ell$ can be encoded as natural numbers, $\mathbb{N}$. Furthermore, because Theorem II-2 applies to all subsystems of $u_0$, its formal mathematical properties are *scale invariant*; that is, the tape for every computation process is quantized but the value of the quantization constant varies from scale to scale. The definition of a system's *scale* and the significance of *scale invariance* are presented formally in §VI.

Despite the foregoing plausibility arguments, the pragmatic criterion of distinguishability is *ad hoc*; it simply "works." Why, ultimately, is overprinting prohibited? Does the finite non-zero length of a tape segment reflect some logical constraint necessary for the operation of a TPM?

Suppose that overprinting were permitted. A computer could print two states, $s_i$

and $s_j = \sim s_i$, on the same tape space — a contradiction.  If overprinting were permitted, no computer could consistently interpret the value of *its own* prior state(s) as input to its current computation.  Since reasoning from a contradiction renders *every* formula provable, if the operations of TPM are inconsistent, all possible states are formally indistinguishable.  The very notion of a distinct state evaporates since $s_i = \sim s_i$.  But that is precisely what happens when the length of a tape segment is either zero or infinite—no two symbols can possibly be distinguished by any logical criterion or by any physical means.  In sum, the equivalence of the logical and physical prohibitions against overprinting is a consequence of Theorem II-2 and the requirement that TPM operate consistently.

The foregoing results are summarized by:

**Theorem IV-1**: *The non-zero finite quantization of the TPM tape is necessary for the consistent operation of $u_0$ or any subsystem thereof.*

Graphically, $u_0$ can be uniquely characterized by a set of points in a three-dimensional tape parameter space that delimits possible values of the arbitrary constants in those laws subject to the constraints: $0 \leq h, G, c \leq \infty$.  This space will be called *constants of nature space* (CON space).  The three dimensions correspond to the properties of quantization, logical cohesion, and maximum tape velocity labeled $h$, $G$, and $c$ in Figure IV-1.  In §VII, I suggest why the values of these tape properties correspond to the familiar physical constants $h$, $G_N$, and $c_{vac}$.

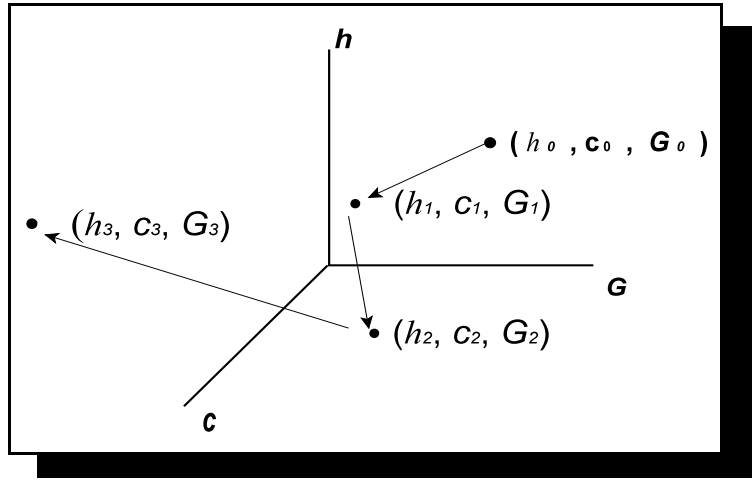

**Figure IV-1**
Constants of Nature Space

Figure IV-1 suggests that it is possible to have multiple decidable universes, $\{u_i\}$, each characterized by particular values of the constants $h_i$, $G_i$, $c_i$. Such is indeed the case. Were there only one decidable universe, $u_o$, then there must exist a rule specifying why CON space is restricted to just the specific values of $h_0$, $G_0$, $c_0$ characterizing $u_o$ and no others. In that case, since $h_0$, $G_0$, and $c_0$ characterize *all* decidable universes, they would be part of the program for *U* — a contradiction since these constants must be algorithmically independent of *U* if *U* is to be consistent. *Hereafter, $u_0$ refers specifically to the decidable universe that we can observe; $u_i$ designates the $i^{th}$ decidable universe formally similar to $u_0$ but characterized by a different trajectory in CON space.* There exist at least $\aleph_0$ decidable universes. Notice that all states of $u_i$ are undecidable relative to $u_j$ and vice versa.

It is important to distinguish between the value of a point in CON space and the address bit in a program occupied by that value. Each arrow in Figure IV-1 corresponds to the addition of a new address bit to the length of the programs being operated upon by a TPM. The addition of a new address bit, that is an arrow in the evolution of a system, adds numbers (states) to the repertory of numbers that can be computed by the TPM. These numbers could not have been computed by TPM prior to the addition of the new address bit.  By contrast, given a fixed number of bits addressable by a TPM, permutation of the values of the addressable bits only moves the destination of each arrow within a given volume of CON space.  Volume in CON space corresponds to the set of all numbers computable by a TPM with programs of length $\leq \mathcal{L}$.  As I shall discuss in §VI, the evolution of the laws of physics has two different connotations: first, the permutation of states computed by a TPM with programs of a fixed length; second, an increase in the number of states computable by a TPM.[b]  When I assert that laws of physics change over time, I refer to changes in the addressable bits of a computer program.  Permutations do not change the information gained as a result of an interaction with a physical system.

In §V it will be argued that the evolution of any given universe, $u_i$, is defined by the set of points tracing a segmented curve in CON space as more and more constants of nature are added to the laws of physics for $u_i$.  Each triplet of constants defines a different scale of computation occurring in $u_i$.  *The state of $u_i$ is the set of all points in CON space associated with $u_i$.*  Notice that logical cohesion requires the state of $u_0$ to comprise all bend points on an entire curve, not just the most recently added point.  It follows that, at any given time, $t_i$, TPM may be performing computations on all scales that have been added to $u_0$ up to $t_i$.  Notice further that the state is defined in terms of tape parameters, not the usual geometrical and mechanical variable of physics.  As is evident from the ability to construct the fundamental Planck scales of geometric length, mass and time from $h$, $G$, and $c$, geomechanical variables grow out of the process of computation; they are not a background stage upon which computation occurs.

The process of adding constants to the laws of physics is best described as a *logical tunneling* from the undecidable states, $\bar{U}$, of $E$ to the decidable states, $\bar{D}$ of $u_0$ (Figure II-1).  Logical tunneling will be symbolized as $\bar{U} \to \bar{D}$ although strictly speaking one should append the subscript $u_0$ to $\bar{D}$ to indicate that the terminus of a tunneling is our universe, $u_0$, and not some other universe, $u_i$. We shall see later that the process is unidirectional; $\bar{D} \rightarrow \bar{U}$ is prohibited on logical grounds. The analogy with quantum tunneling is deliberate. For, just as quantum tunneling is the probabilistic mechanical process by which a particle-wave penetrates a potential energy barrier greater than its kinetic energy, logical tunneling is the probabilistic computational process by which a constant penetrates the logical potential barrier between $\bar{U}$ and $\bar{D}$. The probabilities of quantum mechanics are given by the Schrödinger equation; in §V we shall find that logical tunneling probabilities are given by the algorithmically random Chaitin halting probabilities for programs of length $n$. Translated into physical computers, that is TPM's, in §VI we shall demonstrate that the temperature of $\bar{U}$ is *0°K* whereas the temperature is $\bar{D}$ is always *>0°K*. Thus the analogy with quantum tunneling is complete; a constant, which embodies a set of programs that may be input to a TPM, is penetrating a potential barrier that is greater than its kinetic energy.

---

[b] Later in this section, it will be shown that the logical cohesion prohibits a decrease in the number of states computable by a TPM.

There exists one very important exception to the constraint $0 \leq h, G, c \leq \infty$: namely, a TPM *all* of whose theorems are mutually undecidable or algorithmically random. These theorems are consistent with each other because no one theorem or any collection thereof asserts a property that negates the properties of any other theorem. They are logically independent. No subset of such theorems suffices to prove any other member of the set. In this case, the values of $h, G$ and $c$ are themselves undecidable. Such systems are guaranteed to exist by Gödel's First Theorem and have significant physical consequences for $u_0$. At the end of §V, it will be argued that one can attribute a "temperature" of $0°K$ to such a system. Then, given the incompleteness of $u_0$, the set of algorithmically random states "surrounding" $u_0$ in CON space exerts a (negative) vacuum pressure on $u_0$ that may explain the cosmological expansion of $u_0$.

Before ending this section it is important to point out the while the tape constants, $h, G$ and $c$ discussed here look like the familiar physical constants $h, G_N$ and $c_{vac}$, we have no right to automatically equate the two families of constants. The tape constants specified in this section are logical properties that must characterize the machine tape if a TPM is to avoid overprinting and operate consistently. Only in §VII will I argue that we are justified in equating the tape constants $h, G$ and $c$ with the empirical geomechanical constants $h, G_N$ and $c_{vac}$.

## V. Symmetry-Breaking and Algorithmic Randomness in the Laws of Physics

Theorem IV-1 specifies the necessary condition for the consistent operation of the laws of physics. I now consider sufficient conditions for their consistency.

If the laws of physics are consistent, there must exist at least one state of nature (formula) that is not derivable from those laws. Thus the experimental measurement of one such formula among all the formulae characterizing $u_0$ is sufficient evidence for the consistent operation of the laws of physics governing $u_0$ and every subsystem thereof. In order to qualify as evidence of sufficiency, one must prove that the candidate formula is formally undecidable with respect to any possible laws of physics absent that formula.

Do undecidable formulae exist among the states of $u_0$? Given Theorem II-2, the existence of such formulae is guaranteed by Gödel's First Incompleteness Theorem. (Gödel-1)[35] Thus the observation of undecidable formulae is precisely what one requires for sufficient evidence of the consistency of the laws of physics. The existence of undecidable formulae among the states of $u_0$ implies that the laws of physics are incomplete at any given moment in the history of $u_0$. Gödel's Second Theorem (Gödel-2) asserts that the formalized arithmetic of $(\mathbb{N},+,\times)$ cannot formally prove its own consistency.[36] As one might expect, among those undecidable formulae is the formula corresponding to the statement, "The operation of the laws of physics governing $u_0$ is consistent."[37] But which formula or formulae?

The necessity of tape quantization suggests that the measurement of a specific, non-zero value of the quantization parameter will provide sufficient evidence for the consistent operation of $u_0$ or any subsystem thereof. In order to show that it is sufficient evidence, one must prove that any specific value of the quantization parameter is algorithmically random with respect to all other states of $u_0$ and with respect to the values of all other parameters that could possibly characterize $u_0$. I now claim that symmetry-breaking constants of nature in the laws of physics correspond to the successive bits of Chaitin's halting probability, $\Omega$, for a TPM where:

**Definition V-1**: $\Omega$ is the probability that a computer program, $p_i$, chosen at random, will cause $u_0$ to halt on at least some state $s_i \in \{s_i\} \equiv S$. $S$ is the set of all possible states of $u_0$. At least some $s_i \in S$ are undecidable states.

**Definition V-2**: $\Omega_n$ is defined to be the first $n$ bits of $\Omega$. $\Omega$ and $\Omega_n$ are, respectively, the *total* probabilities that $u_o$ will halt on *any* state and the probability that $u_o$ will halt on at least one state requiring $n$ bits or less to address.

**Definition V-3**: $\Omega_i$ is defined to be the probability that $u_o$ will halt on some state requiring *exactly i* bits to address.

Gregory Chaitin has proved that the individual digits of $\Omega$ are algorithmically irreducible random numbers.[38] Knowledge of any subset of the digits of $\Omega$ provides not the slightest hint as to the value of any other digit. Therefore measurement of the specific value of one or more of those bits of $\Omega$, corresponding to one or more constants of nature for $u_0$, constitutes sufficient evidence for the consistent operation of $u_0$. The specific values of the constants of nature in the laws of physics characterizing our universe, $u_0$, are the result of a random process for which there can be no prior explanation. Their values are simply random facts only determinable by experiment. No theory can explain them. The random process of generating the bits of $\Omega$ will now be shown to correspond to successive symmetry-breakings of the laws of physics.

The *algorithmic entropy*, $K(s^*)$,[c] of a binary sequence, $s^*$, is defined as the length in bits of the shortest program, $p^*$, that computes $s^*$ on a UTM. Because $K$ is defined as the *number of bits* addressing the shortest program, its magnitude is of order $log_2(N_p)$

---

[c] Algorithmic entropy is also called *algorithmic complexity* or *algorithmic information content*.

where $N_p$ is the binary number encoding $p^*$. Formally,

$$K_U(s^*) \equiv \|p_U^*\| \tag{V-1}$$

The symbol $\|s^*\|$ designates the length in bits of a binary string $s^*$.

In terms of $K_U(s^*)$ herewith an alternative definition of algorithmic randomness to that given in §IV:

**Definition V-4**: A sequence $s^*$ is said to be *algorithmically random* if $\|p_U^*\| = \|s^*\|$ to within an additive constant.

The subscript "$U$" in V-1 indicates the machine dependence on $a_U$ of any computation even though any UTM can perform the same computations. The additive constants, $a_U$, constitute the length in bits of the string required to specify the tape parameters defining $u_o$.

For any given physical machine, $U_i$, as $\|s^*\|$ increases, $\|a_U\|$ becomes relatively insignificant. However, if one performs a sequence of computations on different physical machines, the size of the binary sequence $\|\{a_{U_i}\}\|$ coding all the tape parameters can equal $\|s^*\|$. For example, if one performs every calculation on a different machine, then $\|\{a_{U_i}\}\| = \|s^*\|$ to within an additive constant corresponding to the number of bits that define the *machine-independent* universal Turing machine program, $U$. The fact that $\|\{a_{U_i}\}\| = \|s^*\|$ gives rise to a dual representation of randomness: namely, *instrumental* or *mechanical randomness* defined in terms of the number of physical machines performing a computation rather than by the number of steps comprising a program as is the case for *algorithmic randomness*.

In sum, the evolution of a system has equivalent dual representations:

- In the *algorithmic representation* a system's evolution is conceived as a bit-stream corresponding to the tape parameters, $\{a_i\} \equiv \{a_{i,i-1}\}$, of successive TPM's expanding one law of physics containing that constant of nature, $a_i$, into the next.

- In the *mechanical* or *instrumental representation* a system's evolution is conceived as a bit-stream corresponding to the tape parameters $\{s_i\} \equiv \{s_{i,i-1}\}$ of successive TPM's expanding one instrument receptive to that initial condition, $s_i$, into the next.

In the algorithmic representation the free parameters of the evolution are the tape parameters $\{a_{i,i-1}\}$. When $\{a_{i,i-1}\}$ contains the minimum number of elements capable of generating the states $\{s_i\}$, then the elements of $\{a_{i,i-1}\}$ are conventionally called "*constants of nature*." The elements of the set $\{a_{i,i-1}\}$ comprise the points of CON space. In the instrumental representation the free parameters are the states $\{s_{i,i-1}\} \equiv \{s_i\}$. When $\{s_{i,i-1}\}$ contains the minimum number of elements capable of generating the tape parameters $\{a_i\}$, then the elements of $\{s_{i,i-1}\}$ will be called "*initial conditions*." The elements of the set $\{s_{i,i-1}\}$ comprise the points of *initial conditions space* (IC space). Hereafter, I shall consider only the algorithmic representation although the two are completely equivalent and their duality reveals important relationships between the evolution of a system and the evolution of instruments observing that system.

Since $\Omega$ is the probability that at least one program will cause a physical system to halt on some state $s_i \in S$ at some moment in the history of our universe, $\Omega$ is the probability that $u_0$ will occupy all those states that it does occupy during its entire history given the set of states that it might possibly have occupied. The statement appears tautological. Surely, the only possible states that $u_0$ could occupy are those states that it does occupy at some point in its history? Does $\Omega = 1$? No. List all the states that the universe ever does occupy in sequence and map that sequence into the subset $\mathbb{N}_{+,\times} \equiv \{(\mathbb{N},+,\times)\} \in \mathbb{N}$. The mapping is an isomorphism because $\mathbb{N}_{CT} \equiv$ {Church-Turing computable numbers} $\subseteq \{(\mathbb{N},+,\times)\}$ and, per Theorem II-2, the set of all states of $u_0$ equals $\mathbb{N}_{CT}$. But Gödel-1 asserts that $\{(\mathbb{N},+,\times)\} \subset \mathbb{N}$. Hence $\Omega < 1$.

Because all states ever occupied by $u_0$ are members of $\mathbb{N}_{CT,}$ per Theorem II-2 one might suppose that any law of physics has the form

$$s_n = U(s_{n-1}, s_{n-2}, \cdots, s_0) \tag{V-2}$$

where $U$ is the functional form of a UTM in the algorithmic representation of a system. However, such is not the case. V-2 does not in itself determine the precise value of the state $s_n$ for any specific physical system. Define the bit-strings $s_i^* \equiv (s_i, s_{i-1}, \cdots, s_0)$ and $s_{i-1}^* \equiv (s_{i-1}, s_{i-2}, \cdots, s_0)$. Note that $K(s_i^*) \geq K(s_{i-1}^*)$. The algorithmic complexity of a bit-string is only determined to within a machine-dependent additive constant. Therefore the conditional algorithmic complexity of $s_i^*$ given $s_{i-1}^*$ is

$$K(s_i^* \mid s_{i-1}^*) = K(s_{i-1}^*) + (\square a_{U_i} \square - \square a_{U_{i-1}} \square) \tag{V-3}$$

Because the $a_U$ specify the *number of bits* required to describe the hardware parameters of the particular TPM upon which a computation is performed, the values of these parameters are given by $\log^{-1}(a_u)$. Thus V-2 itself only determines the *value* of the state $s_i$ to within a *multiplicative* constant $a_{i,i-1}$ with the magnitude unity when $a_{U_i} - a_{U_{i-1}} = 0$ and $O(\log^{-1}(a_{U_i} - a_{U_{i-1}}))$ when $a_{U_i} - a_{U_{i-1}} \neq 0$. V-2 now becomes

$$s_n = a_{n,n-1} U(s_{n-1}, s_{n-2}, \cdots, s_0) \tag{V-4}$$

V-4 tells us that if a system has evolved through a sequence of states $(s_{n-1}, s_{n-2}, \cdots, s_0)$, then, in the presence of some appropriate physical circumstances specified by $a_{n,n-1}$, the system will evolve into some next state $s_n$. $s_n$ need have no properties in common with any other state in the sequence $(s_{n-1}, s_{n-2}, \cdots, s_0)$. V-4 reveals the evolution of a system through an expanding set of points in CON space illustrated in Figure IV-1.

Notice that there exists some $\{a_n, a_{n-1}, \cdots, a_0\}$ that is the *minimal* set of constants required to compute the evolution of the system. Since $K(s_i^*)$ is only determined to within a machine-dependent additive constant, each element of the minimal set is necessarily incomputable by U from any subset of the set that does not include the designated constant $a_i$. U can compute any number that it is possible to compute but it cannot compute the elements of $\{a_n, a_{n-1}, \cdots, a_0\}$ because, as mentioned above, the number of bits addressing its programs are only determined by the formalism of U to within the machine-dependent constant. It follows that the minimal set $\{a_i\}$ must be algorithmically random. Quite simply there is nothing in U specifying whether a particular computation is being performed on an abacus, an electronic or a quantum computer. The information defining what physical computer is being employed must be added to U ad hoc. This deficiency is revealed by the incomplete definition of $K(s_i^*)$.

The constants, $\{a_{i,i-1}\}$, constitute *symmetry-breaking parameters* because they designate one state from among all possible states of a system to be the next state of the universe, $u_0$. $a_{n,n-1}$ is *symmetry-breaking* because U remains invariant under the substitution of all possible values of $a_{n,n-1}$ in V-4. U (including all prior incorporated $a_{i,i-1}$, for $i<n$) is symmetric with respect to the set of all possible *next* states of a system because the bit string $s_{i-1}^*$ always remains what it is; it is unaffected by the computation of the state $s_i$. Once $a_{n,n-1}$ has been specified, it is incorporated as an axiom in the laws of physics required to compute the state $s_i$ even though, prior to the computation of $s_i$, it was completely independent of those laws. As mentioned earlier, logical cohesion requires that an $a_{i,i-1}$ cannot be removed from the laws of physics even if it is never again employed in the computation of any other state.[39]

What do the constants, $\{a_{i,i-1}\}$, represent? It is useful to expand the recursion V-4 explicitly:

$$s_n = a_{n,n-1} U(a_{n-1,n-2} U(\overbrace{\cdots}^{n-3} a_{2,1} U(a_{1,0} U(s_0)) \overbrace{\cdots}^{n-3})) \tag{V-5}$$

Since both (V-4) and (V-5) assume that *each* $a_{i,i-1} U(a_{i-1,i-2} \cdots a_{1,0})$ is the minimal program to generate $s_i$ from $s_{i-1}$, (V-5) is the minimal program to generate $s_n$ from $s_0$. V-5 considers the evolution of a system as a chronological sequence of states; however, at any given time, $t_n$, logical cohesion requires that the system is performing computations with all programs of *length* or algorithmic complexity $n$ or less steps. Grouping each member of the sequence by program length, $\mathcal{L}$, the *spectrum* of states $s_n^* = (s_n, s_{n-1}, \cdots, s_0)$ has the algorithmic representation:

$$
\begin{aligned}
s_0 &= a_{0,0} \\
s_1 &= a_{1,0} U(s_0) \\
s_2 &= a_{2,1} U(a_{1,0} U(s_0)) \\
&\vdots \\
s_n &= a_{n,n-1} U(a_{n-1,n-2} U(\overbrace{\cdots}^{n-3} a_{2,1} U(a_{1,0} U(s_0)) \overbrace{\cdots}^{n-3}))
\end{aligned}
\qquad \text{(V-6)}
$$

Each $s_i$ corresponds to the halting of a program consisting of $i+1$ steps. $s_0$ is the zero length program, a constant, consisting just of one step: namely the printing of that constant. Any system must contain at least one such constant; otherwise, it is only a vacuous formalism that never computes any particular number or state.

Because $U$ is a universal computer and the $a_{i,i-1}$ are only constants,[40] one can collect the constants together outside the domain of $U$. V-6 becomes

$$
\begin{aligned}
s_0 &= a_{0,0} \\
s_1 &= a_{1,0} U(s_0) \\
s_2 &= a_{2,1} a_{1,0} U(s_0) \\
&\vdots \\
s_n &= a_{n,n-1} a_{n-1,n-2} \overbrace{\cdots}^{n-3} a_{2,1} a_{1,0} U(s_0)
\end{aligned}
\qquad \text{(V-7)}
$$

What are the $a_{i,i-1}$ in V-7? Since V-7 portrays the simultaneous concatenation of possible computational results from the operation of $U$ in its next computation and since $U$ can be programmed to perform *any* formal mathematical operation except the computation of algorithmically random numbers, each $a_{i,i-1}$ constitutes the conditional halting probability that, if U has not halted on a program of *i-1* steps, then it will halt on a program of *i* steps. But I have shown that the $a_{i,i-1}$ are algorithmically random. Hence the $a_{i,i-1}$ correspond to the algorithmically random symmetry-breaking individual digits of Chaitin's halting probability, Ω. Setting $a_i \equiv a_{i,i-1}a_{i-1,i-1} \cdots a_{2,1}a_{1,0}$ we have:

**Theorem V-1**:

$$\begin{aligned} s_0 &= a_0 \\ s_1 &= a_1 U(s_0) = \Omega_1 U(s_0) \\ s_2 &= a_2 U(s_0) = \Omega_2 U(s_0) \\ &\vdots \\ s_n &= a_n U(s_0) = \Omega_n U(s_0) \end{aligned} \qquad \text{(V-8)}$$

with $\sum_{i=1}^{\infty} \Omega_i = \Omega$.

How are halting probabilities related to the familiar constants of nature such as *h*, $G_N$ and $c_{vac}$? Notice that the incomputability of $a_{i,\,i-1}$ implies that the next tape symbol could be either "*0*" or "*1*" for no determinable reason at all. That is only possible if $P$("1"=*property$_i$ is "on"*=TPM halts on a program of length *i*) = $P$("0"=*property$_i$ is "off"*=continue) = ½. Thus the generation of digits on the tape constitutes a Bernoulli ½-process and the relations between different digits for strings with more than one digit must be distributed as the binomial distribution with $P(1) = P(0) =$ ½. The same conclusion follows on information-theoretic grounds. It is not possible to gain more information about the next state of a system than to go from a condition where the next state is incomputable to one where that state is computable.[41] The Bernoulli ½-process provides the maximum amount of information that can be gained from the observation of a process characterized by a two-valued variable. In short, $a_{i,i-1} = \frac{1}{2}$ and $a_i = 2^{-i}$ for all *i*.[42] For an *n*-ary system with *n* primitive symbols in its alphabet rather than the two symbols "*0*" and "*1*", the distribution is multinomial with $P(i)=1/n$ for all *i*. The number of primitive symbols in the alphabet will be called the *symbolic base* of the system.

The finding that all $a_{i,i-1} = \frac{1}{2}$ would appear to contradict my thesis that the constants of nature are random. It does not because the sequence of program lengths for programs that cause U to halt vary randomly as in a perfect coin toss.

In V-6 and V-7 programs are grouped by program length. What is program length, $\mathcal{L}$? By definition it is the algorithmic complexity, $K_U(s_i)$, of the minimal program that computes $s_i$. Its value is $log_2(s_i)$ plus the number of bits required to encode U. But notice in V-8 that the program for $s_i$ is comprised of the number of bits encoding U plus the number of bits encoding $\Omega_i$. If we let $K(s_i) \equiv K_U(s_i) - K(U) = K(\Omega_i)$. Since K is always a positive number, it follows immediately that $K(s_i) = -\log_2(\Omega_i)$. In other words, program length for $s_i$ is the negative logarithm of the halting probability $\Omega_i$ for the program computing $s_i$.

How does $K(s_i)$ manifest itself? How is it measured? Recall that the length of a program is the number of steps executed by U before it halts having produced a result. If we define *time* to be a count of the number of theorems proved or the number computational *results* performed by U, U takes the same "time" to carry out a computation of one step or $n$ steps: namely, one theorem proved regardless of the number of steps required to prove that theorem. This definition of time corresponds to the normal concept. As a result of the motion of its springs and the irreversible pawls of its escapement, a clock proves the theorems whose values are $1, 2, \cdots$, each of which is itself defined to be the cumulative count of the number of theorems proved by the clock. It is essential to point out that there is no way to tell how long each tick takes except by calibrating the duration of each computation by a clock relative to those performed by some other clock. It follows that K is a count of the number of steps carried out by U per computational result achieved by U, that is, per unit time relative to the cosmological clock $u_0$ which ticks no more than once every Planck second. Thus $K(s_i)$ is the *frequency*, $v_i$, of the computation $s_i = \Omega_i U(s_0) = a_i U(s_0)$.

This definition of frequency seems counterintuitive but is not. For example, one does not assert that $n$ pulses of high frequency light take longer to occur than $i$, $i<n$, pulses of low frequency light. Rather, during the unit time in which our clock has performed one computation ("tick"), our observing instrument has registered $n$ pulses. The only difference is that, insofar as $u_0$ is concerned, we have no external clock against which to judge the rate of theorem proving by $u_0$ except with reference to some subsystem of $u_0$ itself. Here time is measured in terms of $U$'s own clock which "ticks" once every time $U$ completes a computation (halts) and produces the result that $U$ uses as input for its next computation.

It is very important not to confuse the frequency of a computation defined above with the halting probability that $U$ will compute $s_i$ as its next result.[43] The common identification of probabilities with the relative frequency of a given event's occurrence within a sample of events is the source of many difficulties in probability theory. The halting probability is simply a primitive concept quantifying the likelihood that $U$ will halt next on a program of algorithmic complexity $K(s_i)$ or frequency $v_i$. Although the frequency of a computation and the probability of that process's occurrence are completely different, they are, nevertheless, intimately related:

**Theorem** $'v_i = K(s_i) = -\log_2(\Omega_i) = -\log_2(2^{-i}) = i$  (V-9)

Like $\Omega_i$, $v_i$ is a pure number; only when calibrated with reference to some standard computer (a clock) does it assume the dimension of reciprocal time.[44]

In V-8 there remains just one constant to consider: namely, $s_0 = a_0$. It is the length of the zero step program. It is just the arbitrary length of the quantum tape segment, $h$, necessary for the consistent operation of the laws of physics. The arbitrariness of $h$ follows from the scale invariance of the laws of physics that is a consequence of the universality of $U$. If we now provide $U$ with the program "•$h$ and $\Omega_i \to -\log_2(\Omega_i)$" ≡ "multiply by the constant $h$ and map $\Omega_i$ into $-\log_2(\Omega_i)$", V-8

becomes

$$s'_0 = h_0$$
$$s'_1 = a_1 U(h_0) = -\log_2(\Omega_1)h_0 = h_0 \upsilon_1$$
$$s'_2 = a_2 U(h_0) = -\log_2(\Omega_2)h_0 = h_0 \upsilon_2 \quad \textbf{(V-10)}$$
$$\vdots$$
$$s'_n = a_n U(h_0) = -\log_2(\Omega_n)h_0 = h_0 \upsilon_n$$

$s'_i$ indicates the state $s_i$ transformed by the above program. If, as I shall argue in §VI, the logical constant $h_0$ corresponds to Planck's constant, then $s'_i \equiv E_i$ : that is, the "energy" of the program generating $s_i$. Hence

**Corollary V-1**: $E = h\upsilon$                                                           **(V-11)**

Corollary V-1 enables one to define the *temperature, T*, of an ensemble of programs as the average energy of the ensemble.

**Definition V-4**: *Provided that the set of all of programs comprising an ensemble is definable, the temperature, T, of the ensemble of programs is $h_0$ multiplied by the average program length of the ensemble.*

Given Definition V-4 one can construct other thermodynamic properties characterizing the ensemble. For example, the thermodynamic *entropy, S*,[45] of the ensemble is just $k_B$ multiplied by the average program length of the ensemble. $k_B$ is Boltzmann's constant expressed in bits. It is beyond the scope of this paper to develop the thermodynamics of computation in any detail except to make the following observation concerning the temperature of an ensemble of undecidable or algorithmically random states.

Consider a system comprised entirely of algorithmically random programs. These are just zero length programs. In fact, from the zeroth entry in V-9, the programs are just arbitrary constants $h_o, h_1, h_2, \cdots, h_i, \cdots$ bearing no algorithmic relation to each other. From V-10 it is evident that the energy of each member of the ensemble is zero and hence the temperature of the entire undecidable ensemble is $T_\mathcal{U} = 0°K$. In other words, the temperature of a set of algorithmically random, mutually undecidable states of a system is $0°K$. Furthermore, the Third Law of Thermodynamics, which is the zeroth entry in V-9, when $k_B$ is substituted for $h_0$ asserts that if $T_\mathcal{U} = 0°K$, then $S_\mathcal{U} = 0 \, ergs/°K$. The zero entropy of the set of undecidable states of a system tells us that the set has no microstates. That is, the set of undecidable states of a system acts like a single state. Such is to be expected from the definition of undecidability. If all members of a set of states are mutually undecidable, there is no algorithm for differentiating between one member of the set of such states from another. Were such not the case, then each undecidable state would be decidable by the rule that distinguishes it.

Given the incompleteness of the laws of physics, the singleton nature of a set of undecidable states is important because it implies that the complete ensemble of all states comprising *E* is *normalizable* relative to any decidable subset of *E*; that is, the ensemble of all states of *E* acts like all the decidable and therefore differentiable states of $u_o$ plus a single program of length zero. It is this characteristic of undecidable states that satisfies the proviso in the definition of temperature. Were such not the case, it would be impossible to calculate a finite value for the temperature or any other functional property of $u_o$. That is, the laws of physics would appear to be algorithmically random even though they are consistent.

Because the proviso is met, the temperature of *E* equals the temperature of its decidable subset, $u_o$. However, *E* consists of the decidable set of states, $u_o$, with temperature $T_{u_0} > 0°K$ plus a single undecidable state $\mathcal{U}$ with temperature $T_\mathcal{U} = 0°K$ Therefore $\mathcal{U}$ exerts a negative pressure or suction on $u_o$ causing $u_o$ to expand in CON space. The presence of $\mathcal{U}$ within *E* is manifest in $u_o$ by the irreversible randomly timed incorporation of the constants, $a_{i,i-1}$, in the mathematical laws governing $u_o$. As a result, the Turing mechanical version of the Second Law of Thermodynamics becomes "The number of arbitrary, symmetry-breaking constants in the laws of physics is an algorithmically random monotonically increasing convex function of time." Alternatively, a symmetry once broken can never be globally restored for $u_o$.

**VI. Genomic Evolution:** *The Growth of Order out of Randomness in the Planck Era*

The randomness of the laws of physics implied by theorem V-1 seems to suggest that an objective search for order in our universe by physics or any other science will fail. Such is not the case.

Consider first the question of objectivity. The following corollary is an immediate consequence of Theorem V-1 since the randomness of the digits of $\Omega$ are sufficient for the consistent operation of $U$.

**Corollary VI-1**: *Given PMP and given elements of $\{a_{i,i-1}\}$ corresponding to digits of $\Omega$, $U$ operates consistently when supplied with any input program $a_{i,i-1}$.*

Provided one can experimentally determine the value of at least one $a_{i,i-1}$, then we know that $U$ can consistently compute an infinite number of states of a system $S = a_{i,i-1} \otimes U$. All but $2^i$ of these states will be repetitions of the $2^i$ states that a law of physics of algorithmic complexity $i$ can compute. The algorithmic randomness of each $a_{i,i-1} \in \{a_{i,i-1}\}$ makes possible all laws of physics even as it makes each law incomputable *in advance* from any other law. Without the existence of random facts — which, by their very randomness, cannot be manipulated by *any* preconceived rationale and must be accepted by *any* being in any specific universe as the "way things are"— objective knowledge could not exist. In this view, consistency and objectivity are two sides of the same coin. Algorithmic randomness makes possible a decision between two alleged outcomes of any logical reasoning. Without Corollary VI-1 science would be impossible.

While Corollary VI-1 establishes the existence of objective laws of physics, it might be the case those laws would be so complex as to be incomprehensible. Indeed, given the randomness of $\{a_{i,i-1}\}$ and given the incompleteness of $(\mathbb{N},+,\times)$, the complete laws of physics for $E$ are incomprehensible; they cannot be expressed in a finite number of symbols related by a finite number of operations. On the other hand, in actuality we do comprehend an extraordinary compass of the physical universe in precise detail ranging from elementary particles to superclusters of galaxies. How is that possible?

The problem of comprehensibility is not just of concern to scientists. Given PMP, if the laws of physics were too complex, elementary particles, let alone biological organisms, could never compute their next state because it would take more than the lifetime of the universe for each particle to process all the interactions among all other particles of the universe. No particle or organism would ever change its state.

Why can we understand the behavior of parts of the physical world without incorporating the properties of every elementary particle in the Universe?[46] Why is the physical world comprehensible in terms of a few relatively simple mathematical relationships? There are two aspects to this problem: first, when is it legitimate to consider the behavior of a system in isolation from other systems comprising $u_0$; second, why can we often comprehend the behavior of relatively complex systems comprising subsystems of very different types or scales? The first aspect concerns the property of *algorithmic locality*; the second concerns the rate of evolution of the laws of physics.

**Definition VI-1**: *A system, S, is algorithmically local if $S = a_{i,i-1} \otimes U$.*

Thus a system is algorithmically local if knowledge of $a_{i,i-1}$ and $U$ is sufficient to decide (predict) all possible states of $S$. Corollaries VI-2 and VI-3 assert that $a_{i,i-1} \otimes U$ is indeed an algorithmically local system.

> **Corollary VI-2**: *Given U, in the constants of nature basis, $S = a_{i,i-1} \otimes U$ is algorithmically local; given I, in the initial conditions basis, $S = s_{i,i-1} \otimes I$ is algorithmically local.*

Because each digit, $a_{i,i-1}$, of $\Omega$ is incomputable from any other, algorithmic locality is guaranteed by the randomness of the digits of $\Omega$ *given knowledge of U*. The second clause of Corollary VI-2 asserts that, when designing an experiment, one does not have to take account of the properties of every other observing instrument in the universe including the interactions of elementary particles that constantly "observe" each other in the course of proceeding from one state to the next.

> **Corollary VI-3**: *Given any $a_{i,i-1}$, all possible states of $S = a_{i,i-1} \otimes U$ are computable by U; given any $s_{i,i-1}$, all possible outcomes of an experiment $S = s_{i,i-1} \otimes I$ are computable by I.*

*If we know $a_{i,i-1}$*, the ability to comprehend any subsystem of $u_0$ depends only on our knowing the Turing-Post programs that define *U*. Corollary VI-3 does not imply that it is possible to construct a finitistic "Theory of Everything" that is applicable for the entire history of $u_0$ because it is conditioned by the proviso "given *any* $a_{i,i-1}$".

Corollary VI-3 suggests that the laws of physics are *scale invariant*, a property sometimes equated with the universality of physical law.

> **Definition VI-2**: *The scale of a system is the number of distinguishable states, $N_{a_{i,i-1}}$, computable by $a_{i,i-1} \otimes U$. Alternatively, the scale of a system is the algorithmic complexity, $K_{a_{i,i-1}}$, of the set of all states computable by $a_{i,i-1} \otimes U$.*

The equivalence of the two versions of Definition VI-2 to within an isomorphism is apparent when we consider that a program for computing a number *is* that number. The program for a number is simply a compact way of writing it. For example, the number $10^{10,000,000}$ takes about 32 million bits to write out (address) longhand; however, the program "multiply the number *10* by itself ten million times" only requires about 38 bits to write out. Nevertheless, so far as *U* is concerned, both the program and the longhand number are the same entity in that each causes *U* to perform the same computation and come to the same result when read by *U* as input on a tape. The isomorphism between the two versions of a system's scale is $K_{a_{i,i-1}} = \log_2 N_{a_{i,i-1}}$.

Because the randomness of the constants of nature (Theorem V-1) guarantees the independence of each $a_{i,i-1}$ from any other, the only component, other than $a_{i,i-1}$, required for computing the states of *any* system, $S_i$, is *U* or *I*. Hence

> **Corollary VI-4**: *The laws of physics are scale invariant; they preserve their form, U, under any transformation $T_{scale}: a_{i,i-1} \to a_{j,j-1}$ over constants of nature space in the algorithmic representation of $u_0$ and they preserve their form ,I, under any transformation $T_{scale}: s_{i,i-1} \to s_{j,j-1}$ over initial conditions space in the instrumental representation of $u_0$.*

Corollary VI-4 asserts that the laws of nature must have the same simple form for small, large, simple or complex systems: namely, *U* or *I* conjoined to the idiosyncratic bits addressing the particular scale $a_{i,i-1}$. Under the transformation $T_{scale}: a_{i,i-1} \to a_{j,j-1}$, only the number of bits required to express the constants defining each scale is increased from $K_{a_{i,i-1}}$ to $K_{a_{j,j-1}}$. Hence $T_{scale} = K_{a_{j,j-1}} - K_{a_{i,i-1}}$ or, equivalently, $T_{scale} = \log_2 \left( \frac{N_{a_{j,j-1}}}{N_{a_{i,i-1}}} \right)$.

The fraction within the parentheses is the ratio of the number of macrostates computable by *U* from the macroscale constant $a_{j,j-1}$ to the number of microstates computable by *U* from the microscale constant $a_{i,i-1}$. In other words, it is the quantity $W_{j,i}$ of classical thermodynamics for CON space. The subscript indices *j* and *i*, *j>i*, are appended here to indicate that *W* is determined relative to the scales defined by $a_{j,j-1}$ and $a_{i,i-1}$ whereas in thermodynamic treatments the algorithmic complexity of the units (e.g. atoms in a gas) is implicitly assumed to be unity.[47] Hence $T_{scale} = \log_2 W_{ji}$. If we multiply $T_{scale}$ by $k_B$, one obtains Boltzmann's equation, $S = k_B \log_2 W_{ji}$, for entropy expressed in bits. Because constants of nature are always added to the laws of physics but never deleted therefrom, the entropy of $u_0$ increases over time; however, in CON space unlike the phase spaces of statistical mechanics, the increase of entropy of $u_0$ is absolutely irreversible. There is no Poincaré recursion in CON space. As stated at the conclusion of §V, in Turing Mechanics, a symmetry once broken can never be globally restored for $u_0$.

In sum, because the constants of nature are algorithmically random, it is always legitimate to consider the behavior of a system in isolation from the rest of $u_0$, *provided that the laws of physics are formulated as the Turing-Post rules U or I and provided that consideration is restricted to those phenomena deducible from just a single constant $a_{i,i-1}$*. However, experimental conditions often span many scales. For example, the observation of microscopic "quantum" systems by "classical scale instruments.

As a brief aside in light of the foregoing discussion of $T_{scale}$, it is appropriate to consider the connection between Planck's equation $E = h\nu$ and Boltzmann's equation $S = k_B \log W$ expressed in bits. $\nu$ equals program length, $\mathcal{L}$; $W_{ji}$ is the halting probability, $\Omega_{j|i}$, for programs of scale *j* relative to programs of scale *i*. But, as we have seen earlier, $\mathcal{L} = -\log \Omega_i$. This suggests that Planck's equation and Boltzmann's equation are simply two different representations of the same process: namely the halting of *U* on a program of length $\mathcal{L}$. Planck's equation expresses the localized mechanical energy manifest by the halting of *U*; Boltzmann's equation expresses the amount of information manifest by the halting of *U* as related to the probability that *U* does halt on $\mathcal{L}$. It looks as though Planck's equation and Boltzmann's equation mirror the particle-wave duality of quantum mechanics except that the probability waves in the Born interpretation of quantum mechanics become entropy waves in Turing Mechanics.[48] In this context, there is one very significant difference between QM and TM: namely, in QM the Hamiltonian is invariant under unitary transformations whereas in TM, owing to the incompleteness of $u_0$, unitarity or probability conservation is not strictly true unless one includes undecidable states among the possible values of an observable.

The conflation of scales in many experimental situations raises the second aspect of the comprehensibility of nature and the evolution of natural laws. Why can we comprehend, perhaps not completely, complex systems spanning multiple scales? Part of the answer concerns the extent to which short (simple) programs provide a good approximation to the full model of a multiscale system. The bits for *U* or *I* remain the same but, for a multiscale system, the arbitrary constant to which *U* or *I* is applied consists of a long string of bits with "1"'s and "0"'s scattered throughout. The distribution of halting probabilities for programs of length *i*, $\Omega_i$, suggests why such systems are comprehensible: namely, $\Omega_i = 2^{-i}$. That is, the shortest program embedded within the multiscale arbitrary constant provides the most likely explanation for the halting of *U* or *I* on any given experimental outcome. Therefore,

> **Corollary VI-5**: *P*(*U* halts) *is maximal for the states of a system corresponding to the first halting digit of $\Omega$ for $u_0$. Furthermore, P(U halts$|\Omega_i$ )is conditionally maximal for the states of a system corresponding to the first subsequent halting digit of $\Omega$ for $u_0$ after the $i^{th}$ digit of $\Omega$.*[49]

Corollary VI-5 explains why perturbation techniques work well in physics: the first term in the series approximation to a function corresponds to the shortest program in the set of programs that we assert to model a physical system. The corollary is the algorithmic rationale for Occam's razor.

If the model proposed herein is correct, within its first Planck second $u_0$, was governed by no laws and possessed no mass-energy. Today it is characterized by marvelously simple and seemingly timeless, universal laws governing the interplay of immense quantities of matter and energy. How did our universe get from there to here? Some evolutionary process must be at work. The remainder of this section considers a simple mechanism that may possibly account for the evolution of the universe from its completely random beginning to its current highly ordered condition.

How rapidly do the laws of physics evolve? How does the regularity that we observe throughout the universe come about if arbitrary constants can be added to the laws of nature at any moment and on any scale from the smallest particle to the largest supercluster of galaxies? I shall now argue that the probability of such changes is extremely small and growing smaller with each computation by $u_o$. Given that the set of algorithmically random numbers dwarfs the set of compressible sequences of numbers, this result would seem virtually impossible. However, for any TPM consisting of $U$ coupled to a finite set of constants of nature, the set of all algorithmically random numbers always appears to have only one element. Hence the probability of a new constant's appearance as the result of any particular computation is generally very small. Why?

Before proceeding further it is necessary to define what one means by the evolution of the cosmos. Specifically, one must distinguish between two types of evolution: *symmetric* or *permutational* evolution and *symmetry-breaking* or *genomic* evolution.

> **Definition VI-3**: *The set of programs of any given fixed length will be called the genome for the scale of program length $\mathcal{L}$.*[50]

Thus a genome consists of a *fixed* number of bits, each of which may assume different values drawn from a *fixed* set of *letters* (formally, primitive symbols) comprising the genome's alphabet. The number of symbols in a genome's alphabet will be called the genome's *symbolic base*. Any specific sequence of values occupying a genome's address bits constitutes a *genotype* or what I have elsewhere called a *program*.

> **Definition VI-4**: *Symmetric or permutational evolution is the interchange of different letter values in one or more of the address bits comprising a genome.*

Symmetric evolution is recognizable as the normal dynamical evolution of a system except that it is an inherently random process.[51] The maximum rate of symmetric evolution is one permutation/ computation-sec, $t_p$. As discussed in §IV, the value of $t_p$ is set by the values the fundamental tape constants $h$, $G$, and $c$. The process must be random because, were it not, then the number of address bits could be reduced by some functional combination of the compressible bits until the values of all bits comprising the genome are algorithmically random. The randomness of symmetric evolution not only agrees with quantum mechanics but also suggests that the randomness of QM itself arises from the requirement that the mathematics of QM be consistent.

It is classically asserted that a computer proceeds from the calculation of one definite number to another. For example, 2+1=3 might proceed to 3+2=5, etc. However, no physical computer always produces exactly the same number when a given computation is repeated. Rather, a distribution of numbers is produced that is determined by the distribution of halting probabilities associated with the programs in a computer's repertory.[52] Corollary VI-5 explains why the classical result is so frequently observed: the distribution of halting probabilities is dominated by the halting probability for the shortest halting program in the repertory.

Symmetric evolution is a conservative process. By definition, the conserved quantity is program length $\mathscr{L}$. Since program length is the frequency of a process and equals $E/h$, conservation of energy follows automatically. Per Nöether's Theorem, every conserved quantity in physics is associated with a symmetry. In this case the symmetry is the group formed by permutating values or primitive symbols among the different address bits of a program's genotype. Thus each genotype gives rise to a symmetry group of particles and/or states of the system $TPM_{\mathscr{L}}$ and the symmetric evolution of a system is just the exchange of one member of the group for another.

It is to be observed that the symmetric evolution of $TPM_{\mathscr{L}}$, if it occurs in $u_0$, is complete in that every element of the group as well as combinations thereof can be computed by the operation of $U$ on the set of constants $\{a_{i,i-1}\}$ associated with $TPM_{\mathscr{L}}$. When we assert that the laws of nature are unchanging, we mean that evolution of $u_0$ is only symmetric. Given the incompleteness of (ℕ,+,×), that assertion is manifestly ruled out even though symmetric processes may, in some circumstances, provide highly accurate approximations to the evolution of $u_0$ within a given scale era.

To incorporate the incompleteness of (ℕ,+,×) in the laws of physics it is necessary to consider asymmetric or genomic evolution.

**Definition VI-5**: *Genomic evolution is the addition of one or more address bits to the address bits comprising a genome.*

Since the deletion of address bits from the genome for $u_0$ is prohibited by logical cohesion, genomic evolution is also *asymmetric* and will be referred to as such on occasion. The converse of Nöether's Theorem requires that genomic evolution not be a conservative process since it is asymmetric. Each step or *mutation* in the genomic evolution of a system increases the length of the programs in the repertory of the TPM and therefore necessarily increases the system's total energy by $h\mathcal{L} = h\nu$. Here $\mathcal{L}$ is the total number of addressable bits *after* a mutation. Genomic evolution is a metamathematical process giving rise to programs none of which could be obtained from the symmetric permutation of any of the ancestor genome(s). The definition does not rule out the genomic evolution of a zero bit program to a single or multiple bit program such as occurred within the first Planck second of $u_0$. Indeed, instead of conceiving genomic evolution as the addition of an additional address bit to an existing multiple bit program, an alternative is to view genomic evolution as the addition of a multiple bit program to a zero bit program. From this point of view the addition of each new set of constants to the laws of physics is a symmetry-breaking process. Combining these two views, the first of which conceives of the evolution of $u_0$ to be the algorithmically random addition of new constants to the laws of physics for $u_0$ (logical tunneling) and the second of which conceives the evolution of $u_0$ to be a symmetry-breaking of *E*, one must conclude that a symmetry-breaking of *E* gives rise to a new symmetry in $u_0$.

To calculate the rate of genomic evolution in the course of any computation by a TPM, it is sufficient to notice that one is not interested in the probability that a program with specific heretofore incomputable values in each address bit will be incorporated in the next computation.  Rather one seeks the probability that a new address bit will be added to the bits addressable by $TPM_{\ell-1}$ regardless of the values occupying any of the addressable bits.  It follows that the sample space consists of the set of possible address bits.  Since the halting probability is defined in terms of halting on programs of length *1*, *2*, ⋯ plus the lengths of all programs of undecidable length, the sample space has one element for each program length.  I have shown that the set of all undecidable programs has only one element since there is no rule by which to distinguish between undecidable programs of different lengths.  Thus, at any given time in the history of $u_0$, the sample space contains one element for each of the currently addressable bits plus one element for the set of programs whose length is undecidable.  Given these conditions, the following stochastic process models the genomic evolution of $u_0$ from an algorithmically erratic condition within its first Planck second to the mathematically effective condition that we observe today.

Consider an urn filled with one ball colored black to represent its undecidable state. (Figure VI-1) The urn will *always* contain exactly one black (undecidable) ball because $(\mathbb{N},+,\times)$ is always incomplete but it will only contain one black ball because there is no rule within $(\mathbb{N},+,\times)$ by which to distinguish between undecidable states.

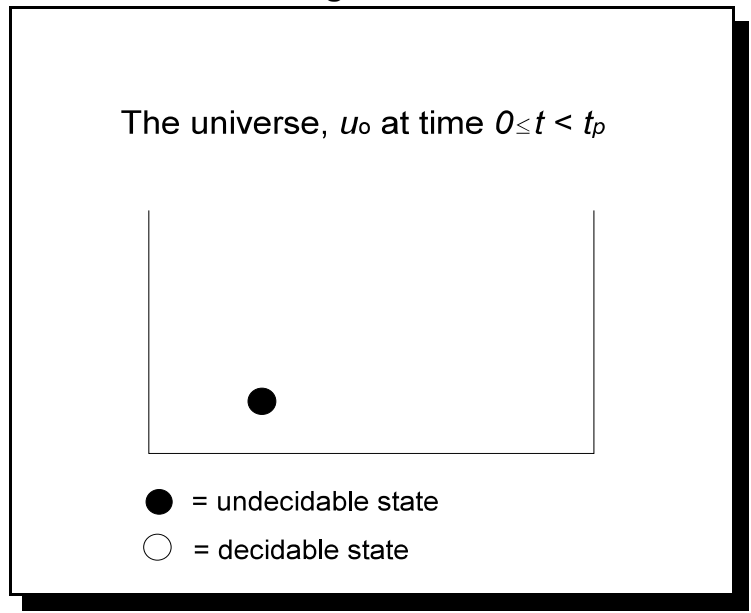

**Figure VI-1**

The universe, $u_o$ at time $0 \leq t < t_p$

● = undecidable state
○ = decidable state

Now imagine an invisible (metamathematical) hand reaching into the urn, picking out a ball.  A disembodied voice inspects it and announces "the ball is black." (Figure VI-2)  In so doing, the ball is transformed into a white ball, corresponding to a decidable state simply because "the voice" has asserted that this particular ball is in some decidable state called "black" by "the voice."  The ball changes color (from black to white) because its announced state ("black) is no longer undecidable.  This color change corresponds to the process of logical tunneling.  Notice that, during the interval in which the hand and disembodied voice are inspecting the ball and announcing "the ball is black," the state of whatever balls may occupy the urn is undecidable.  There is no way, at that moment, to tell what the state of any ball in the urn might be.  As shown in Figures VI-1 through VI-3, the process of withdrawing a ball from the urn, inspecting it, and then returning it to the urn corresponds to the three steps in a computation: namely, reading some input bit from a tape, transforming the value of that bit, and writing the transformed value of the bit onto the tape.

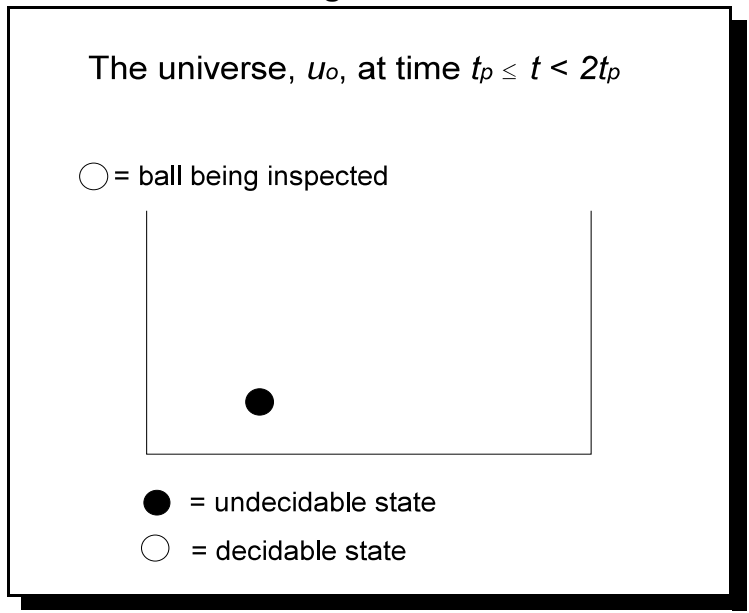

Figure VI-2

The "hand" now returns the ball to the urn and the resulting state of $u_0$ is given by Figure VI-3.

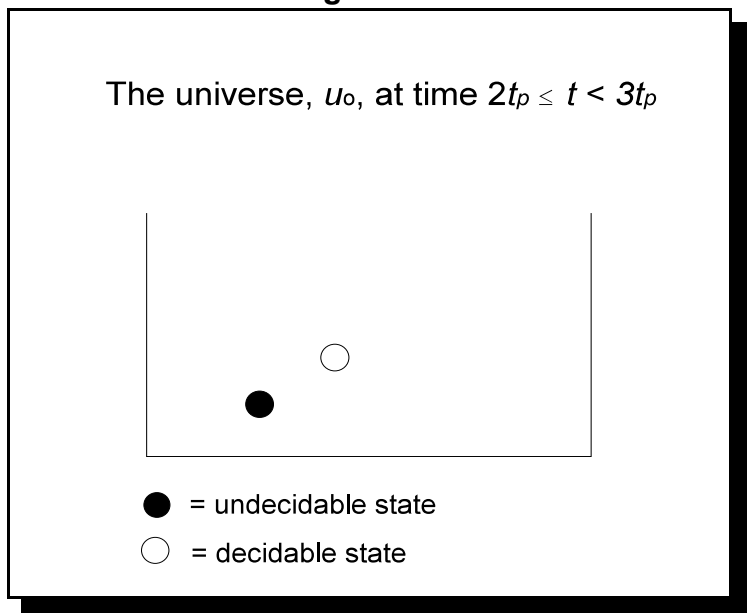

Figure VI-3

Why does the inspected ball remain "white" when returned to the urn? Shouldn't it revert to a black undecidable ball when it is replaced in the urn? Owing to the asymmetry of genomic evolution, the inspected ball remains white. An undecidable ball (black) may be converted into a decidable ball for "no reason at all," but a decidable (white) ball can never revert to an undecidable (black) ball owing to the requirement for logical cohesion.

The "hand" now reaches into the urn a second time but on this occasion it only has a 50% chance of drawing out a black ball. After many draws, the state of the urn looks like Figure VI-4:

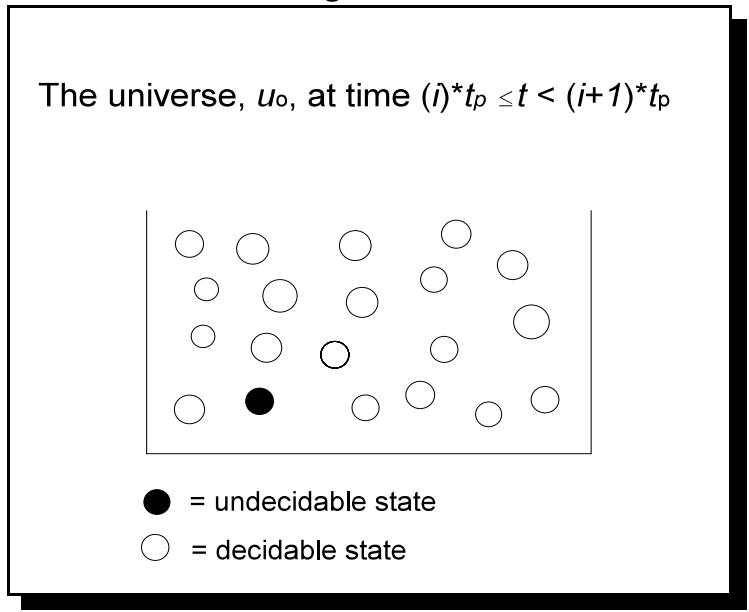

**Figure VI-4**

The universe, $u_o$, at time $(i)*t_p \leq t < (i+1)*t_p$

● = undecidable state
○ = decidable state

After $n$ inspections the probability, $P_U$, of picking an undecidable state and encountering a new arbitrary constant in the laws of nature characterizing a previously undecidable property of $u_0$ is $P_u = \dfrac{1}{t}$ ; the probability, $P_D$, of picking a decidable state of a type encountered previously is $P_D = \dfrac{t}{t+1} = 1 - P_u$. $t$ is the number of inspections that have occurred.

At the present time, when $t \approx 3 \times 10^{60} t_p$ where $t_p$ is the Planck second, the chance of encountering an undecidable state is minuscule. Thus the laws of physics appear universal and unchanging to us. Even at the end of the Grand Unification era, when $t \approx 10^8 t_p$, the laws of physics would have appeared remarkably stable. The chance of then encountering an undecidable state would only be $\approx 1/10^8$. However, such would not be the case had one been around to observe the universe within the first thousand Planck seconds. Within that period, algorithmic randomness of the laws of nature would have been quite noticeable.[53]

If $N(t)$ is the number of arbitrary constants required for the symmetric but only approximately complete model of $u_0$ at time, $t$, then

$$N(t) = \sum_{i=1}^{t} \frac{1}{i} \qquad \text{(VI-1)}$$

Graphically, that is

**Figure VI-5**

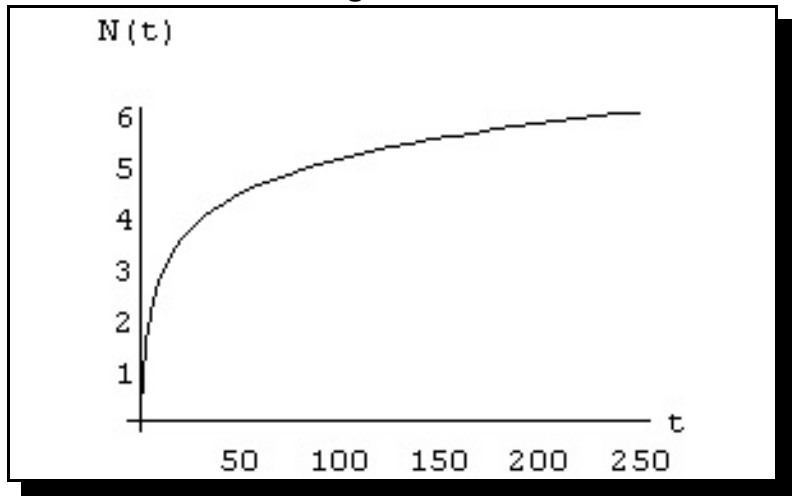

In Figure VI-5 $N(t) \to 0$, and $dN(t)/dt \to \infty$ as $t \to 0$ while $dN(t)/dt \to 0$ when $t$ and $N(t) \to \infty$. Thus, within the first Planck second of $u_0$, the laws of physics are materially undefined in that they contain no material constants and are completely unstable;

however, the laws become asymptotically stable over time even though the number of constants in those laws becomes arbitrarily large.

At the present time, if our universe is typical of other universes that have survived at least as long as ours, the laws of physics for $u_0$ should contain

$$N(t_p) \approx \sum_{t_p=1}^{10^{63}} \frac{1}{t_p} \approx \ln 10^{63} + \gamma \approx 146 \qquad \text{(VI-2)}$$

arbitrary constants. $\gamma$ is Euler's constant. Do the symmetric laws of physics for our time really contain 146 arbitrary constants? If the "Standard model" of strong, weak, and electromagnetic forces contains just nineteen free parameters[54], and even that many makes many physicists uneasy,[55] how is one to interpret the prediction of equation VI-2 that the laws of physics now contain 146 constants of nature?

Notice, as a consequence of Corollary VI-3, that properties of the scale created by a symmetry-breaking step in the genomic evolution of $u_0$ are fixed for all time. Thus, when calculating the number of algorithmically random constants required in a theory purporting to model $u_0$ up to and including the scale created by a symmetry-breaking, one must determine the number of constants required *at the time the symmetry-breaking took place*. The Hadron era, when the forces modeled by the Standard Model first appeared as distinct forces began when $u_0$ was approximately $10^8$ Planck seconds ($10^{-35}$ seconds) old[56]. Hence the number of arbitrary constants in the Standard Model should be

$$N(t_p) \approx \sum_{t_p=1}^{10^8} \frac{1}{t_p} \approx \ln 10^8 + \gamma \approx 18.99 \approx 19 \qquad \text{(VI-3)}$$

The agreement between VI-3 and the Standard Model which contains nineteen free parameters plus the triplet $h$, $G$, and $c$ specific to the Planck era may be only a coincidence; nevertheless, it is surely merits further investigation.

The question remains what we are to make of the extra *127* free parameters apparently required by VI-2 for a "complete" symmetric model of our universe in the current era of its history. At this point I can only speculate that these extra parameters may define complex structures including life that have arisen since the onset of the Hadron era.

To summarize, the genomic evolution of $u_0$ will never be completed; however the pace of its evolution declines rapidly as $u_0$ ages while the compass of physical laws increases exponentially. Mathematically,

$$\lim_{t \to \infty} \frac{\log_2 N(t)}{N(t)} = 0 \qquad \text{(VI-4)}$$

*N(t)* is the number of states that can be formally generated by a TPM at time *t*. $\log_2 N(t)$ is the number of bits required to address all the states of the system as well as the number of free constants in TPM's program at time *t*.

In this section, I have modeled the genomic evolution of $u_0$ as the never ending accumulation of algorithmically random constants in the laws of physics applicable to $u_0$ through the process of logical tunneling. The process begins (Figure VI-1) with $u_0$ in an undecidable state characterized by no laws at all for $t_{u_0} \leq t_p$. Given this characterization of $u_0$ for $t_{u_0} \leq t_p$ and given that $T_{\mathcal{U}} = 0°K$ for any ensemble of undecidable states[d], we have

**Corollary VI-6**: $T_{u_0} = T_{\mathcal{U}} = 0°K$ for $t \leq t_p$.

It should be recognized, of course, that the appearance of a finite duration of time characterizing the "creation" of $u_0$ is a retrospective view from our current position in the history of $u_0$. Were one actually present at that moment, there would have been no duration at all until after the first "theorem" was proved by *U* when the Planck scale constant tunneled to join $U$.[57]

---

[d] §V ending.

## VII. $h$, $G_N$, $c_{vac}$ and Other Evidence for the Consistent Evolution of our Universe

In §IV three logical constants, designated $h$, *G*, and *c*, were shown to define material properties of a Turing machine tape that are necessary for the consistent operation of any TPM to avoid overprinting on the tape. The properties correspond to *quantization*, *logical coherence*, and *maximum tape velocity* respectively. If the Principle of Mathematical Physics (PMP) is valid, then, at very least, observations of our physical universe, $u_0$, must yield evidence that these three properties are manifest in the known laws of physics characterizing $u_0$. If we can find constants of nature with those properties, then that empirical finding would provide evidence for the existence of the tape properties necessary for the consistent operation of mathematical laws in $u_0$.

Clearly, the empirically measured constants $h$, $G_N$, and $c_{vac}$ manifest the three required properties. $h$ reveals the maximum divisibility or quantization of $u_0$ although quantization occurs in action space and only incidentally in geometric space-time. The operations of TPM are quantized in action owing to our ability to perform *rotations* in the axiom-theorem space of a formal system. A formal system generates a set of formulae (*axiom-theorem space*). Some small number of these formulae are termed axioms. From them all other members of the set, called theorems, can be proved by the application of *U* to the axioms. However, any particular choice of axioms constitutes an arbitrary *representation* of the formal system. Many equivalents are possible and may be constructed by replacing any given axiom with some appropriate combination of the theorems in the initial representation. The process of replacing axioms by combinations of theorems is analogous to the rotation of basis vectors in a vector space. Hence the use of the term "rotations" in a logical context. While the formulae of axiom-theorem space remain the same regardless of which representation one employs, the number of steps required to prove formula *B* from formula *A* varies greatly depending on the particular representation one employs. In one representation the computation of *B* from *A* will take many steps or units of computational time; in another the computation will require only a few steps. For example, in one representation the computation might be performed with no intermediate steps by a single program of length $\mathcal{L}$ whereas in another it might be performed in four steps by four programs of length $\mathcal{L}_1, \mathcal{L}_2, \mathcal{L}_3$ and $\mathcal{L}_4$. It is evident that conservation of program length between *B* and *A*, or, equivalently, conservation of energy between state *B* and state *A* asserting that $\pounds = \sum_{t=1}^{4} \pounds_t$, is only possible if the action product $\pounds \cdot (t=1) = (\sum_{t=1}^{4} \pounds_t) \cdot (t=4)$. For any given axiom-theorem space, there exists some minimal subset of formulae that will generate all other elements of the space. This subset comprises the smallest or orthogonal basis set of axioms generating the space. The elements of that subset are algorithmically incompressible. The cardinality of the minimal subset is, by definition, the algorithmic complexity, *K*, of the space. In the minimal representation the computation of *B* from *A* requires the maximal number of steps required to perform the computation with each step comprising the minimal program length possible within the given action-theorem space. The minimal action product $(\mathcal{L}_o) \cdot (t=1) = h_o$ constitutes the quantization constant for a given axiom-theorem space. For $u_0$, this constant is Planck's constant, $h$.

The fact that the fundamental logical domain of physical computation is defined in action-space not geometric space has been suggested by others. Drawing on the work of Gödel and Chaitin, Reginald Cahill, Christopher Klinger and their colleagues have demonstrated that algorithmically incompressible interactions (self-referential noise) between pseudo-particles modeled on Leibniz's monads[58] give rise to enduring relationships between these monads similar to the discussion in §VI. In their model, these relationships constitute topological defects embedded in the *derived* three-dimensional geometric space on which the events governed by classical and quantum physics are played out.[59,60,61,62,63,64] The logical connection between consistency and quantization in action seems to have been anticipated by Turing. In a footnote to his 1936 paper on the nature of computation,[65] Turing suggests that symbols must be distinguishable in action space. It appears Turing came very close to discovering that Planck's quantum of action may arise from the need to distinguish symbols in a physical universe operating as a computer. Unfortunately, since he was considering only the design of idealized mathematical machines and not their application as an essential basis for physical theories, he did not draw the conclusion that quantization is necessary for the consistent operation of a TPM in $u_0$.

The empirically constant velocity of light in *vacuo*, $c_{vac}$, clearly corresponds to the logical tape parameter $c$. Both specify the maximum velocity at which information can be propagated through space in the case of light and along the machine tape in the case of a TPM.

$G_N$, Newton's gravitational coupling constant, quantifies the cohesion of space-time by determining the force or curvature of space that must be surmounted to "escape" the gravitational field between one body and another each with given masses separated by a spatial distance $r$. The mass of each body corresponds, via Einstein's equation, $E = mc^2$ to a given energy. That energy in turn corresponds to a program of length $\mathcal{L}=v =E/h$. But the cohesion of space-time is exactly what the logical constant $G$ expresses with respect to the Turing machine tape: namely, the conditions under which the tape will tear and a TPM can no longer access the results of its past computations given the specific values of $h$ and $c$ characterizing $u_0$ for a given program of length $\mathcal{L}$. Notice that gravitation is the only purely attractive force between normal matter[66] as expected if it embodies logical cohesion. Also, like logical cohesion, gravitation has an infinite effective range subject to physical conditions under which space "tears." Examples of such tearing may occur at the event horizon of a black hole or at the event horizon of $u_0$. The event horizon of $u_0$ constitutes the logical boundary between decidable theorems and undecidable formulae. For these reasons I conclude that $G_N$ embodies the logical cohesion of $u_0$. The extreme weakness of gravity relative to other fundamental forces is an accidental consequence of the algorithmic randomness of the constants of nature in $u_0$. In another universe, $u_i$, $G_N$ could be much stronger or even weaker.

With regard to the tearing of space-time or of the machine tape, it deserves mention that were some segment(s) of the machine tape to be torn from others under extreme conditions, the "floating" segment would be unable to "perform" any computation; the segment would itself constitute an algorithmically random constant unless, somehow, it became attached as a new constant of nature to some universe, $u_i$. That possibility is determined by the halting probability for the length of the floating segment. On this basis it appears that laws of physics would not operate within the event horizon of a black hole. The possibility of a wormhole connection from the black hole to another universe $u_i$ or a reconnection to some other region of $u_0$ would be inversely proportional to the length of the floating segment. The larger a black hole is the less likely it is that its interior will connect to another universe. In the limit case for the totality of $u_0$ itself, it is virtually inconceivable, but not impossible, for $u_0$ to fuse with another universe $u_i$.[e] Only in this statistical sense can we say that $u_0$ is isolated from all other decidable universes, $\{u_i\}$.[67]

In sum, the empirical appearance of the three constants $h$, $G_N$ and $c_{vac}$ — the only constants that can be combined to form units of mass, length and time defining the conditions of our universe in the Planck era—suggests that the necessary condition for the consistent operation of the laws of physics is met at the most elementary level of our universe. Therefore:

> **Conclusion VII-1**: *The constants $h$, $G_N$, and $c_{vac}$ with the properties of the logical constants $h$, G, and c in the laws of physics constitute necessary evidence for the consistent operation of U and I in $u_0$.*

I have designated the foregoing a "conclusion" because, unlike the previous theorems and corollaries, it can never be formally proved by *U* given the constants characterizing $u_0$. Rather, one can only assert that available evidence suggests the legitimacy of equating the logical constants $h$, $G$ and $c$ with the physical constants $h$, $G_N$, and $c_{vac}$. If that suggestion is accepted, then their presence in the laws of physics provides necessary evidence for the consistent operation of those laws in $u_0$.

---

[e] Such is also a consequence of equation VI-4

What evidence would be sufficient to demonstrate the consistent operation of $U$ in $u_0$? In §V, it was asserted that, if $U$ is to operate consistently, then $U$ cannot prove every possible formula.[68] Specifically, $U$ must not be able to formally derive specific *values* of the tape parameters differentiating one physical realization of $U$ from another. In other words, one must be able to show that observed values of the constants of nature in the laws of physics are algorithmically random.

On the face of it, considering the extent of CON space (Figure IV-1), the formulae $h = 6.63 \times 10^{-27}$ erg-sec, $G_N = 6.67 \times 10^{-8}$ gm$^{-1}$-cm$^3$-sec$^{-2}$ and $c_{vac} = 3.00 \times 10^{10}$ cm-sec$^{-2}$ seem to be contingent. Contingency implies that their truth can only be determined experimentally.[69] There is nothing in the known formal relationships of physics, such as $E = h\nu$ that in any way determines the value of $h$ given that $E$ and $\nu$ must be allowed to range over all possible values. As remarked in connection with Figure IV-1, if the constants characterizing $u_0$ were not contingent that implies $u_0$ is the only decidable universe. But were there only one decidable universe, $u_o$, then there must exist a rule specifying why CON space is restricted to just the specific values of $h_0, G_0, c_0$ characterizing $u_o$ and no others. Then, since $h_0, G_0, c_0$ would characterize *all* decidable universes, they would be part of the program for $U$ — a contradiction because the constants must be algorithmically independent of $U$ if $U$ is to be consistent.

On the other hand, the constellation of values of the constants of nature seems to exhibit a near miraculous fine tuning that gives rise to the delicate conditions under which life could arise to evolve the human mind capable of looking out on the universe and becoming conscious of the laws of nature that themselves gave rise the human mind (the anthropic argument). To some this fine tuning seems impossible without the presence of a designer or some convergent evolutionary process[70] determining the constants' values. However, the anthropic argument by design fails, even if that design were a convergent stochastic process. First, the complex structures of our universe including life are only unusual and the necessity for a designer only arises if one assumes that our universe is the only decidable universe. As we have seen, such cannot be the case if the laws of physics are to operate consistently. Many other universes could, and certainly have, developed structures at least as complex as those in our universe but life in those universes would look very different than the life with which we are familiar. Further, suppose that the observed values of $h$, $G_N$, and $c_{vac}$ were not contingent and could somehow be computed *a priori* by *U*. Then one must be able to determine experimentally some other values of these tape parameters, $h'$, $G'$, and $c'$, that could not be computed by *U* if *U* is to operate consistently. In that case, $h'$, $G'$, and $c'$ would be merely renamed versions of $h$, $G_N$, and $c_{vac}$. In other words, the designer or convergent process would operate inconsistently were the constants of nature not algorithmically random at some scale of reality — again a contradiction.

Another argument in favor of the algorithmic randomness of $h$, $G_N$, and $c_{vac}$ is suggested by the remarks above. Suppose one compiled all known laws of physics together and then eliminated any redundancies by reducing the entries in the catalogue to the minimum possible by including only those laws that can be proved to be algorithmically independent of each other. If the total number of variables and constants in the catalogue, $n_{v \cup c}$, were greater than the number of equations, $n_e$, relating them, then $n_{v \cup c} - n_e$ of the constants must be algorithmically independent and random relative to the known laws of physics. One would conclude that all available evidence suggests the laws of physics operate consistently. So far as we know, such is indeed the case but hope persists among some who seek a "theory of everything" that we will eventually be able to state all laws of nature in a finite set of equations containing no inexplicable constants. Of course, were such a hope realized, it would shatter those very laws by rendering them inconsistent. Thus this aspiration is an anathema to physics but physics must, nevertheless, always seek it — if only to show that it is unattainable. Once again, as was the case for the previous two arguments, were arbitrary constants to be eliminated from the laws of physics a contradiction would result since the derivations of the very laws that contained no such constants would themselves be inconsistent. Physics would have to rebuild all its findings from the ground up to try to discover new laws containing arbitrary constants. That attempt might be unsuccessful since the procedure outlined in this paragraph is ultimately founded upon the belief that nature is consistent. Still the history of physics is replete with surprises. In the present state of our understanding, it is ultimately only a hope that nature does not harbor the supreme deception: namely, that our belief in the rationality of the Universe is illusion.

I now consider two physical arguments in favor of the randomness of the tape parameters. First, the particular values of $h$, $G_N$ and $c_{vac}$ that we employ or that a physical system employs in its computations are only determined to within a certain dispersion that is ultimately dependent on $h$. I argued in §IV that the dispersion is necessary to avoid inconsistency through overprinting on the machine tape. As noted by Turing,[71] overprinting can only be avoided by limiting the number of symbols in the alphabet of a language to a finite number of symbols. The finite alphabet constrains the number of digits by which the dispersion in a measurement is limited since each successive digit defines an algorithmically independent program when fed into a TPM. Now consider an arbitrarily large system of interacting TPM's. Their measurements of the dispersion will only converge on an average value with convergent estimates of the dispersion of the average if the measurement programs of all TPM's are mutually consistent. But that consistency itself requires the tape parameters of at least one TPM to be algorithmically random otherwise the system would operate inconsistently. In other words, if one could determine *a priori* the values of the tape parameters there would be no basis for a convergence by all TPM's on the average values of the tape parameters and their dispersion. Thus the experimental fact that the measurements of all observers converge on an average value of a constant provides evidence for the consistent operation of $u_0$ and also evidence for the algorithmic randomness of the constants of nature.

Second, algorithmically random constants in the mathematical laws of physics imply the existence of undecidable states in the complete Universe, $E$. A uniquely quantum phenomenon may provide direct evidence for undecidable states in $E$ and therefore indirect evidence for the presence of random constants in those physical laws. Decoherence arises from the interaction of the coherent wave function of a quantum computer with its environment. Decoherence is what makes things happen. If a wave function never decohered to the distribution of single values of an observable that appear in our detectors, we would never observe anything.

Consider now our universe as a quantum computer with a wave function described by the Wheeler-DeWitt equation. If the universe described by this equation were complete, how could its wave function collapse or decohere to produce the $u_0$ that we witness? Since decoherence requires the interaction of a wave function with an environment, the wave function of our universe or of the set of all decidable universes must somehow interact with an "environment"—whatever that environment might be. The interaction cannot constitute an algorithmically compressible process because the set of decidable universes, by definition, encompasses all states that can be computed given the specific constants of nature characterizing each universe at any given time in its history. On this basis, I suggest the phenomenon of decoherence provides evidence that our universe, describable by and operating according to consistent mathematical laws, is incomplete. In §V one process was proposed by which the intervention of $E$ in $u_0$ may occur: namely, the symmetry-breaking addition of algorithmically random constants to the laws of physics (logical tunneling).

In sum, the foregoing analysis strongly suggests but does not formally prove the following.

**Conclusion VII-2**: *The observed finite non-zero values of $h$, $G_N$ and $c_{vac}$ are algorithmically random and therefore their specific values in the laws of physics constitute sufficient evidence for the consistent operation of $U = I$ in $u_o$.*

As was the case for Conclusion VII-1, I designate the foregoing to be a conclusion because none of the arguments advanced heretofore can be formalized as a proof of the conclusion and the arguments in favor of Conclusion VII-2 could be refuted by future experiments. Furthermore, all the arguments ultimately assume the very consistency of the laws of physics that they seek to demonstrate. Therefore, at best, they only provide strong indicators of the consistent operation of $u_0$ and of the randomness of the physical constants differentiating $u_0$ from all other decidable universes.

In light of the foregoing limitations, one must ask whether there is any test that could provide evidence for the algorithmic randomness of the constants of nature which does not assume the consistency of $u_0$ as part of its design? A seemingly insurmountable obstacle is the fact that there exists no algorithm capable of definitively deciding whether or not a given finite string of binary digits drawn from an infinite binary sequence is random in the Chaitin sense defined earlier.[72] A way around this obstacle may be to employ Per Martin-Löf's measure theoretic definition of *effective randomness*. To determine whether a string is effectively random one subjects a string to sequential tests for non-randomness. If the string fails all tests (that is, no regularity is detected), then the string is deemed to be *Martin-Löf random*.[73] Chaitin has shown that Martin-Löf randomness is equivalent to *Chaitin* (*algorithmic*) *randomness* in the sense that, if a string is Martin-Löf random, it is Chaitin random and vice versa.[74] The existence of a test by which to determine whether the constants of nature are Martin-Löf random is an open question to be considered in a future paper.

Before ending this section the perspective of Turing Mechanics on two contemporary issues in cosmology and physics deserves mention. The first concerns the dark energy (DE) and the cold dark matter (CDM) that together are estimated to constitute at least 95% of the total mass-energy of our universe. Per Corollary VI-2 (algorithmic locality), the read-write head (language acceptor) of a TPM only scans one segment of the tape at a time while performing a computation and writes the content of the segment into the computer's memory. In addition to the bits storing $U$ whose contents remain unchanged from one computation to the next, the memory contains a data register comprised of exactly the number of bits equal to program length appropriate to the scale at which the TPM is operating. The size of the data register equals the scale size and no more or less in order to ensure algorithmic locality.[75] In the case of a Planck scale computer, the data register comprises just one bit; as argued at the end of §VI, we expect the standard model to contain *19* free parameters in addition to the three fundamental constants $h$, $G$ and $c$ which remain unchanged. Thus we should expect the data register of a hadron computer to comprise *19* bits.

   Working from the data stored in its memory, the TPM performs a computation and writes the result on the tape giving rise to the phenomena that make up $u_0$. So far the computation process is thermodynamically reversible. The TPM now reads another segment to begin the next computation. However, before it can write the new data in its memory, the TPM must erase from memory the data that provided input to the prior computation. Charles Bennett has shown that energy is dissipated when a computer's memory is erased.[76] Where does this dissipated energy reside in $u_0$? It cannot escape from $u_0$ because its density and temperature are not great enough to drive a computation let alone tear the tape. The answer is that it resides on the filled tape trailing from the back end of the computer. At this point it is necessary to recall that the halting probability, Ω, for $U$ is only defined for self-delimiting programs. Self-delimitation requires that all programs in the repertory defined by a constant of nature be *prefix-free*.[77] That is, no program bitstring can be the prefix of another program string.[78] It can be shown that a prefix-free Turing machine must move in only one direction along the machine tape.[79] In other words, once a result has been written on the tape, the result is forever inaccessible to the TPM. Bennett's dissipated erasure energy seems to derive ultimately from the requirement that the set of programs available to a computer comprises a prefix-free language. My equation of the dissipated erasure energy with the energy of past program results written on the trailing tape suggests an interpretation of gravity. Since *G* constitutes the coupling constant that assures the machine tape can never be torn, I suggest that *G* defines the coupling between the present state of *U* and its past history. Thus, gravity may be the geometrical manifestation of constraints placed by the past history of our universe upon its present degrees of freedom. In 1995[80] I argued that the history of our universe since its beginning is embodied in the erased mass-energy dissipated by Planck and higher scale computers. Therein I estimated that the erased mass-energy may constitute 89% of the total mass-energy of our universe and proposed that the dissipated erasure energy is a candidate for the "missing matter" whose gravitational effect was well known but whose origin was unexplained by contemporary cosmological theory. Since that time, the comparison of redshifts and luminosities of distant Type 1a supernovae with those of nearby Type 1a supernovae as well as the power spectrum of the cosmic background radiation has lead to the conclusion[81,82] that expansion of $u_0$ is accelerating. It is believed that the acceleration is driven by the vacuum energy embodied in a positive cosmological constant, $\lambda$. In light of this discovery, my earlier result must be reinterpreted. It is now seen to refer the proportion of CDM to the total *matter content* of $u_0$ rather than to the proportion relative to the total mass-energy of $u_0$ as I originally supposed. After correcting for a downward revision in the age of $u_0$ employed in my prior work from *1.7x10$^{10}$* to the best current estimate of *1.4x10$^{10}$* years, we obtain cold erasure matter as 87.5% and normal computationally active matter as 12.5% of the matter content of $u_0$ respectively. These calculations agree with best astronomical estimates of the matter content of our universe: (29±4)% CDM; (4±1)% baryons and ~0.3% neutrinos.[83] Hence, from those estimates, one obtains 87% CDM/total matter; 13% (baryons+neutrinos)/total matter. What is the cold dark matter? In Turing Mechanics, it is simply the past states of all the currently existing baryons, neutrinos and non-neutrino leptons.[84]

To conclude these remarks on the cosmological missing mass-energy one further question deserves an attempt at an answer: namely, where does $u_0$ get the endless supply of blank tape upon which to write the results of its computations? There is no way that $u_0$ can recycle old tape and write new results on the freed-up space because that would involve further erasure dissipation of energy which then has to be dumped on some other "backup" tape held in some other part of the universe.[85] The answer is that blank tape is endlessly supplied from the undecidable formulae, Ū (Figure II-1) that complete $E$ beyond $u_0$ and beyond all decidable universes, $\{u_i\}$. As new constants of nature logically tunnel into $u_0$ from Ū, they generate an exponentially expanding tape space upon which $u_0$ can write the results of its computations. The "volume" of the tape space, that is the number of possible state values that can be computed by $u_0$ scales as $2^{n_t}$ where $n_t$ is the number of constants in the laws of nature at time $t$. However, if Equation VI-2 relating the number of constants to the age of $u_0$ is correct, then it should be possible to compute the deceleration parameter, $q_0(t)<0$ for cosmological expansion. I leave this topic for a future paper; however, these considerations suggest that the expansion of tape space resulting from the logical tunneling Ū →D̄ may be the source of the "dark energy" currently invoked to explain cosmological acceleration. Dark energy is estimated to constitute two-thirds of the total mass-energy of our universe.[86]

The second contemporary issue concerns the claim of quantum mechanics and its successors, quantum field theory and string theory, to be the fundamental style of model by which to conceptualize physical reality. This issue includes the problem of resolving the relationship between large and small-scale systems. Turing Mechanics agrees with many results of quantum mechanics and of the classical mechanics of relativity theory within their respective domains. The concurrence between quantum and Turing Mechanics is required if Turing Mechanics is to satisfy Bohr's Correspondence Principle; however, Turing Mechanics applies to every scale per Corollary VI-4 (scale invariance). Scale invariance implies that any formal distinction between macroscopic and microscopic systems is unsustainable and that, when the laws of physics for macroscopic systems are cast in algorithmic form, macroscopic quantum effects should reveal themselves.

Unlike its agreement with quantum mechanics, Turing Mechanics is in fundamental logical disagreement with classical continuum mechanics.

**Corollary VII-1**: *The operation of a physical system according to continuous mathematical laws is inconsistent.*

Corollary VII-1 is a direct consequence of the prohibition against tape overprinting discussed in §IV. The corollary asserts that if a physical system were to operate according to the classical continuum mechanics of Newton and of Einstein, then its operation would be inconsistent. Thus the disagreement between classical and quantum mechanics goes deeper far than the experimental failure of classical physics in the microscopic realm of Planck quanta. While quantum mechanics could constitute both a conceptual model of the physical world and of the processes by which it operates, classical physics could never do so. Classical continuum physics constitutes a stunning conceptual model that can never correspond to the operations of physical reality if those operations are to be consistent.

**VIII. Conclusion:** *Physics in the Random Universe*

In order to understand why physical reality is quantized, it was suggested herein that one must first answer the question posed by Eugene Wigner: How can one account for "the unreasonable effectiveness of mathematics in the natural sciences?"[87]

In response it was proposed, *as an hypothesis to be empirically tested* (PMP and Theorem II-2), that nature is a mathematical machine in the sense defined by Church and Turing. The finite non-zero value of Planck's quantum of action, $h$, of Newton's constant of gravity, $G_N$, and of the velocity of light in *vacuo*, $c_{vac}$, provide evidence necessary for the consistent operation of objective mathematical laws governing our universe (Conclusion VII-1). Consistency requires the presence of algorithmically random, symmetry-breaking constants in the laws of physics of any universe. At any given time, $t_i$, these constants correspond to the first *n* digits, $\Omega_n$, of Chaitin's Halting Probability, $\Omega$ for our universe. Thus the specific values of these constants for our universe can only be determined by experiment. It was argued (Conclusion VII-2) that observed specific values of the physical constants *h*, $G_N$ and $c_{vac}$ must be members of an algorithmically random set of constants in the laws of physics. Therefore their observed values and those of all other constants of nature provide sufficient evidence for the consistent operation of $u_0$. It was suggested that the Martin-Löf measure of effective randomness may provide the means by which to rigorously demonstrate the algorithmic randomness of the constants of nature. However, in keeping with Gödel's Second Theorem, it may turn out that such a demonstration is intrinsically impossible.

      Theorem V-1 paints a paradoxical picture of our attempts to understand the laws of physics and the mathematical operations of physical systems. On the one hand, the laws of physics are alleged to constitute random facts; on the other hand, the randomness of these facts (the constants in the laws) is necessary and sufficient for their consistent and, therefore, lawful operation. In addition, the randomness of physical laws is necessary and sufficient for algorithmic locality (Corollary V-2), local completeness (Corollary VI-3), scale invariance (Corollary V-4) and simplicity (Corollary VI-5). The random constants are not *dei ex machina*. Precisely because they are random, they cannot be employed to willfully paper over ignorance of the physical universe. Quite the opposite, the randomness of these constants and therefore the randomness of the laws of physics themselves ensures the objectivity of our knowledge because their values are beyond all computation. They can only be discovered by experiment. They are brute facts that no being can change. All the computing power in our universe and even in all decidable universes cannot ordain their values. Come when they may for no reason at all, our universe, $u_0$, must simply await

the arrival of the next constant of nature from the complete Universe *E*. The Universe is its own explanation.

Should the findings of this paper be sustained under scrutiny by others, then the following conclusion would appear inescapable: one of the most perplexing scientific discoveries of the twentieth century, Heisenberg's Uncertainty Principle, while seeming to limit our knowledge of the physical world, may actually be required for any knowledge of the order in nature. We have seen that the finite non-zero value of $h$ appears to be necessary for the consistent operation of the laws of physics. Some physicist, I believe it to be John Barrow but can't be certain, has written that the name "Uncertainty Principle" is a misnomer. It should be called the "Certainty Principle" for the following reason. In classical Newtonian physics, almost all orbits are unstable. The slightest perturbation in initial conditions or from external influences can catapult a system into wildly chaotic behavior for which no prediction is worth more than a German mark in the 1930's. Not so the atoms of quantum mechanics. Thanks to the Uncertainty Principle and the non-zero constant of action $h$, the electron remains bound to the proton in an hydrogen atom for almost all time.[88] Herein I have suggested that the influence of the Uncertainty Principle and the quantum of action runs far deeper than the mere stability of atoms. Without them, there would be no laws of nature because the operation of any universe would be inconsistent.

The view that order and randomness in nature are intimately related is not new to physics. It is the foundation of quantum theory. What is new is the position that randomness is *necessary* for any order in the physical universe. Others have come to the same conclusion: notably Cahill, Klinger and their colleagues in the work cited earlier. The role played by monads in their work is analogous to the role played by meaningless tape symbols in this paper; the random generation of persisting *n*-ary relationships between monads in their work corresponds to the spontaneous incorporation of symmetry-breaking constants in the laws of physics herein.

The view that there may exist more than one universe, indeed an infinite number of universes, each governed by its own laws of physics, is also not new.[89,90] However, Turing Mechanics is controversial. Does it constitute a scientific theory? The question is paramount because the theory implies that the complete Universe consists almost entirely of undecidable states beyond the compass of *any* rational or algorithmically compressible program.

To qualify as a scientific hypothesis,

- A theory must be objective. It must constitute a formal mathematical theory whose symbols are independent of varying interpretation by one investigator or the next.

- It must be consistent. In order to be consistent, the theory must include as an independent axiom at least one formula (symbol-coded fact of nature) that can only be determined on the basis of experimental evidence.

- It must be refutable.[91] The theory and its derived theorems must contain at least one formula which distinguishes it from any other theory.

- It must not be vacuous. The refutable formula must be shown to correspond to some observed state of a physical system.

Turing Mechanics satisfies all four criteria. Since it is a formal theory, its theorems and predictions are objective because $U$ and $I$ are universal computers. The constants in the laws of physics are objective because they are beyond manipulation and because they are formally incomputable except when incorporated as *ad hoc* axioms in those laws. Second, because Turing Mechanics contains algorithmically random constants, it is consistent. Third, since these constants are algorithmically random, at least one formula in any specific realization of the theory is necessarily distinct from all other formulae generated by all other physical realizations of Turing Mechanics. Thus the theory is refutable. The theory is not vacuous. It is a fact that we do exist in that we measure the values of $h$, $c_{vac}$ and $G_N$. In sum, Turing Mechanics constitutes a scientific theory.

It may be objected that the Principle of Mathematical Physics and Theorem II-2 conflate human procedures for computation with the dynamical evolution of physical realty. The latter, so it is alleged, is certainly different than the action of any human being doing sums. This challenge raises the infamous mind-body problem of Descartes.[92] How are our minds and thoughts connected to the physical world? We are, after all, part of the world and the existence of those thoughts can only be explained if that fact is taken into account. In this paper I have attempted to surmount the mind-body problem by hypothesizing that human procedures for computation and the dynamical evolution of a physical system are, indeed, the same. Thus I account for the effectiveness of mathematics in the physical sciences. It is important to notice, however, that the principle only asserts that human computation and physical evolution have the same form. In all other respects they are as irreducibly different as counting on your fingers is different from the operation of the personal computer upon which I am writing: your fingers and this PC are each characterized by different constants of nature. Whatever its virtues or limitations, the Principle of Mathematical Physics is only an hypothesis. It is only viable if it works.

In this paper four examples were provided to support the thesis that the universe can be conceived as a consistently operating mathematical machine, TPM. First, the observed presence of algorithmically random constants in the laws of nature implies that TPM operates consistently. Second, the phenomenon of decoherence for $u_0$ as a whole seems to predicate the existence of an environment for $u_0$. That environment can only be the undecidable states that complete the Universe, $E$. Third, the mass-energy of erased states created as a TPM clears its memory from one computation to the next was calculated and closely approximates the best astronomical estimates of the CDM fraction of the matter content of $u_0$. Fourth, logical tunneling of constants from Ū →D̄ in Turing Mechanics could be the source of the "dark energy" driving the accelerating cosmological expansion of $u_0$.

      The constellation of the constants of nature that the anthropic principle purports to explain is extraordinary. Unpredictable in advance, in hindsight the finely tuned values of these constants have made possible the exquisite relations that give rise to life on earth and the practice of science itself. But the truly unfathomable beauty of our situation is not those relations. It is that, were the laws of physics not random, then our universe could never come into being and flower as it has. Herein I have suggested that our circumstance can only be accounted for if we posit an heretofore unrecognized category of existence: namely undecidable states of our universe. Since Gödel, undecidable truths have been recognized as a proper category of mathematical existence. If the laws of physics can always be expressed mathematically, it seems difficult to avoid the conclusion that they inhabit physical reality as well.

Acknowledgments: *I am deeply grateful to Dr. Job L. Dittberner, Prof. Ali Eskandarian, and Dr. Anthony B. Wolbarst for their critical assistance. This work was supported by Aerovest, Inc.*

**Notes**

16. Sérgio Volchan takes issue with this point when he asserts:

> "Even more problematic are some allegedly deep connections to randomness in the *physical* world...(The connection) seems to presuppose that the 'universe' can somehow be identified with a kind of big universal Turing machine putting out bits for us to decode. Besides being strongly anthropomorphic, this is an extreme simplification of nature and is at odds with the picture presented to us by the natural sciences."

However, in the end he concedes that "The question is whether or not our description of (physical processes) bears any relevance to the natural processes themselves." Volchan, S.B. 2002. "What is a Random Sequence?" *American Mathematical Monthly*, 109 #1 (1/02) 61f.

  One of the strengths of the Principle of Mathematical Physics is precisely its anthropomorphism because only if we can show that physical processes have the same form as mental processes can we break down the deadly mind-body dichotomy and come to terms with the fact that human beings and their thoughts are part of the physical universe. Furthermore, the Principle is far richer than any specific branch of mathematics or physics because it incorporates as an integral component all the undecidable truths that are essential to complete both mathematics and the physical universe. While it is true that the notion of effective computation is necessarily restricted to discrete processes, that limitation is no hindrance because the success of quantum mechanics strongly suggests that physical reality is discrete. Indeed, at the end of §VII I argue that any physical system operating according to continuous mathematics would operate inconsistently.

  As to the relevance of the theory put forward here, I would emphasize that the Principle of Mathematical Physics is an hypothesis to be tested against experiment. I leave it up to the reader to decide whether the evidence given herein or by others who may choose to follow this line of inquiry does, in fact, justify its acceptance.

17. Deutsch, D. 1997. *The Fabric of Reality.* Penguin Books. New York. (esp. chap. 9.)

18. Dewdney, A.K. 1993. *The New Turing Omnibus*. Computer Science Press (W. H. Freeman and Company), New York, NY:211-214.

19. Casti, J.L. 1992. *op. cit.*:288.

20. Landauer, R. 1985. "Computation and Physics: Wheeler's Meaning Circuit?" in Zurek, W.H. *et. al.* ed. *Between Quantum and Cosmos, Studies and Essays in Honor of John Archibald Wheeler*. Princeton University Press, Princeton, NJ:568.

21. Church, A. 1936. *op. cit.*

constant for its proof.  Therefore the constant is still decidable even though it is never employed by the system.  In sum, the complexity of programs provable by a computer, as measured by the number of arbitrary constants in its repertory of internal states, always grows.

      The property of logical cohesion explains why mathematical and physical truths appear to be eternal verities independent of our willful actions even as we recognize that both mathematics and physics progress historically.  These truths are, indeed, independent of our will because the underlying constants of mathematics and physics are algorithmically random; however, they are not eternal verities.  They have been added to our mathematics and to the laws of physics over time.  There was a time when some mathematical truths and some laws of physics did not exist in that they were not a formal consequence of any mathematical or physical system *in operation at that time*.  Where did these constants "come from" to be added over time to mathematics and to the laws of physics?  They came *for no reason at all* from the endless stock of algorithmically random constants contained in *E* that completes mathematics and the laws of physics.

      It is remarkable that the unidirectional "arrow" of time may arise from the requirement that mathematics be logically coherent and, hence, from the asymmetry of adding but never subtracting new constants to the laws of nature.

35. In fact, Gödel demonstrated metamathematically that there exist at least $\aleph_0$ undecidable formulae that can be enumerated by successive Gödel number mappings. Gödel numbers are defined by the isomorphism $\Gamma : \mathbb{N} \rightarrow \mathbb{N}$ encoding every theorem of arithmetic as an element of $(\mathbb{N},+,\times)$ which itself is a subset of $\mathbb{N}$.

36. See the following sources concerning Gödel's theorems and Gödel numbering: Nagel, E. & Newman, J.R. 1960. *op. cit.*; Casti, J. L. 1992. *op. cit.*:308-332; Rucker, R. 1982. *Infinity and the Mind*. Birkhäuser, Boston, Basle, Stuttgart:267-292; Gödel, K. 1934. *op. cit.*

37. The statement is Gödel's Second Incompleteness Theorem.  When the statement is itself Gödel number coded as a natural number, that number becomes a constant of nature asserting that the laws of physics operate consistently.  From Gödel-2 it follows that this Gödel number must be determined empirically.

38. Chaitin, G.J. 1998. *op. cit.*  See also Beltrami, E. 1999.  *What is Random?* Copernicus (Springer-Verlag) New York, NY.  esp. chapter 4 for a discussion of algorithmic randomness and $\Omega$.

39. Formally, a symmetry once broken can never be restored for $u_0$ although it may be restored locally for a subsystem of $u_0$ (at the cost of large increase in the entropy of $u_0$ ). This result is the generalized version of the Second Law of Thermodynamics for Turing Mechanics.

40. Possible non-commutivity of the elements of a sequence arises from logical cohesion not from the anything in the value of the constants themselves. Hence one can bring the constants outside the parentheses as in V-7.

41. As stated earlier, I assume that all programs generating a sequence of states are minimal programs.

42. An important proviso here is that all programs be *self-delimiting*. See Chaitin, G.J. 1998. *op. cit.*12f. This condition is equivalent to the stipulation that Equation V-7 applies to CON space.

43. The difference between frequency as employed in probability theory and frequency as employed in computation theory is apparent from the following: In probability theory when we speak of the (relative) frequency of an event *a* we mean that in a set of *y* events *a* occurs $x_a$ times and it does *not* occur $y - x_a$ times. In computation theory, per the definition given in the text, we speak of an event's frequency when the event *a* recurs *x* times in a given unit time. In the latter case there is no reference at all to *a*'s non occurrence. In sum, frequency in probability and frequency in computation are very different even though they are functionally related as shown in Equation V-9.

The origin of the confusion between the two concepts of frequency derives from the fact that the halting probability for a long program is less for a long program than for a small program. Hence the belief that short programs take less time to carry out than long programs. Because *U* halts less frequently on long programs than short programs a long program seems to take longer to perform than a short program.

44. GivenTheorem V-2, how can one account for high frequency processes that appear to characterize the computations of quantum scale systems? The answer is as follows. The concept of frequency employed in Equation V-9 is always defined *relative to each computer's own clock* (*computations*). Thus if I am observing a hadron system and my clock completes one computation/second whereas the hadron's clock completes a computation once every $10^{-35}$ seconds, then during one second of my time the hadron system executes a program of apparent length $\mathcal{L} \approx 10^{35}$ steps relative to my clock even though, as we shall find in §VI, the maximum length program upon which a hadron can operate relative to its own clock is estimated to be approximately *19* bits long.

It should be noted that the maximum program length or frequency capacity of a volume of tape space of length $ct_{i,j}$ is proportional to $2^t$. $t_{i,j}$ is the number of theorems that system *i* (for example, a hadron) can compute in the time it takes system *j* (for example, my clock) to compute one theorem. When compared with the halting probability $\Omega_i = 2^{-t}$ for programs of length *t*, one obtains the program length frequency power spectrum of *1/f* noise that characterizes fractal systems. We should expect such to be the case given the scale invariance of the laws of physics demonstrated in §VI.

45. As opposed to algorithmic entropy.  Thermodynamic entropy is measured in units of ergs/$^0$K whereas algorithmic entropy is logarithm of the dimensionless number of steps in a program.

46. Davies, P.C.W. 1990. "Why is the Physical World so Comprehensible?" in Zurek, W.H. ed. *Complexity, Entropy and the Physics of Information*. Addison-Wesley Publishing Company, Reading, Ma. 64f.

47. In other words, the atoms' internal degrees of freedom are not considered in the computation of an ensemble's entropy.

48.  E.T. Jaynes has come to a similar conclusion on information theoretic grounds regarding the relationship between the Planck and Boltzmann equations.  See Jaynes, E.T. 1990.  "Probability in Quantum Theory." in Zurek, W.H. ed. *Complexity, Entropy and the Physics of Information*. Addison-Wesley Publishing Company, Reading, Ma. 381-403 and Jaynes, E.T. 1957.  "Information Theory and Statistical Mechanics." *Phys. Rev.*106, #4, 620-630.

49. Corollary VI-5 is not original to this paper but has been pointed out by others: notably, Beltrami, E.  1999. *op.cit.*115.

50. Notice that, because energy is defined as the observable produced when the program length, $\mathscr{L}$, is multiplied by $h$, each scale is characterized by a fundamental energy level $h\mathscr{L}$.  Such is exactly what we would expect from quantum mechanics.

51. As always I assume that the laws of nature are stated in irreducible form; that is, in the algorithmic basis where the number of constants of nature is the minimum possible and all constants are algorithmically random or orthogonal in CON space.  The practical difficulty is that one can never prove that a given experimental situation defines an orthogonal set of initial conditions.

52. In this view, genomes of length *i* correspond to the set of all Feynman paths of $\leq i$ steps.

     An immediate consequence of Gödel's First Theorem is that the almost all elements of the set of Feynman paths consist of an undecidable number of steps.  Only the singleton property of the set of paths of undecidable length gives rise to a normalizable distribution with $\sum p_i \approx 1$ of Feynman paths of length *i*.

53. On the other hand, owing to the scale invariance of physical laws, the magnitude of the "Planck's constant" defining each scale increases with the magnitude of the scale.  Then the "age" of the universe for macroscopic systems with large $h$ is relatively young even now.  For macroscopic systems, the appearance of undecidable states should be quite noticeable.  However, since these undecidable states are not properties of the physical universe tied to elementary particles, the fundamental logical significance of undecidable states may not be recognized for what they are.  For example, genomic mutations in biological systems constitute the addition of arbitrary constants to a biological genome.  Biological properties expressed by the enlarged genome are undecidable with respect to the original shorter genome.

      The concatenation of genetic segments as a process of logical tunneling suggests a possible explanation for so-called "junk" genes.  Given the asymmetry of genomic evolution, so-called "junk" genes may constitute the biological mechanism by which its is ensured that biological systems will never retrace their evolutionary history to produce maladapted organisms.

54. Greene, B. 1999:142 & 389f.

55. See, for example, Lee Smolin:

> "No theory with twenty parameters that can be freely chosen can be considered to be a fundamental theory of anything.  What is clearly missing are some additional principles that set the values of these parameters."  (Smolin, L. 1997. *The Life of the Cosmos*. Oxford University Press. Oxford, Eng. UK:50.)

Smolin is not alone in his viewpoint.  Dirac and Eddington, among others, believed that the laws of physics could be formulated without reference to any empirically determined constants.  Nevertheless, these concerns may be misplaced.  The constants of nature cannot computed from combinations of a smaller number of other parameters; they constitute independent axioms defining the material properties of $u_0$.  Their independence and their frequency is a consequence of the algorithmic randomness required for the consistent operation of *U* in $u_0$.

56. Davies, P.C.W. 1982. *The Accidental Universe*. Cambridge University Press, New York, NY:29.

57.  Notice that the program for *U* is just one of the constants in Ū during $t \leq t_p$.  Thus the tunneling that formed the first TPM in $u_0$ is simply concatenation of two constants each possessed of equal temperature: namely, $T = 0^\circ K$.  Hence the tunneling occurs with probability *1* in "no time at all."

58. Leibniz, G.W.  *ca.* 1710.  "Monadology" in *Leibniz: Discourse on Metaphysics; Correspondence with Arnaud; and Monadology.* 1957. trans. by George R. Montgomery.  Open Court Publishing Company. LaSalle, Ill.  251-272.

59. Cahill, R.T. & Klinger, C.M. 1996.  *op. cit.*

60. Cahill, R.T. & Klinger, C.M. 1997. "Bootstrap Universe from Self-Referential Noise." arXiv:gr-qc/9708013.

61. Cahill, R.T. & Klinger, C.T. 1998. "Self-Referential Noise and the Synthesis of Three-Dimensional Space." arXiv:gr-qc/9812083 v2.

62. Cahill, R.T. & Klinger, C.T. 1999. "Self-Referential Noise as a Fundamental Aspect of Reality." Contributed to 2$^d$ International Conference on *Unsolved Problems of Noise*, Adelaide. Also arXiv:gr-qc/9905082.

63. Cahill, R.T., Klinger, C.T. & Kitto, K. 2000. "Process Physics: Modeling Reality as Self-Organizing Information." arXiv:gr-gc/0009023.

64. Cahill, R.T. 2001. "Process Physics: Inertia, Gravity, and the Quantum." arXiv:gr-gc/0110117 v1.

65. Turing, A.M. 1936. *op. cit.*:135 (footnote).

66. I exclude here the so-called "dark energy" which will be discussed in §VII. Dark energy clearly has a repulsive gravitational effect.

67. From these results and Corollary VI-6, it follows that the temperature of the interior of a black hole is *0°K*. This circumstance explains why, classically, a black hole is black and only absorbs matter from $u_0$. However, because the reconnection between the interior and some universe, including $u_0$, is dependent on the halting probability, $\Omega_i$ for a program of length $\mathcal{L}$, the temperature *at the event horizon* is *>0* and inversely proportional to the mass of the black hole as a consequence of the mass-energy-program length relationship. Given the random distribution of digits of the halting probability, black hole radiation ought to constitute a black body spectrum as suggested by Stephen Hawking.

68. For example, see Nagel and Newman's demonstration of the absolute consistency of the propositional calculus. Nagel, E. & Newman, J.R. 1960. *op. cit.* 54. This statement is the converse of the fact that one can prove any formula from a contradiction.

69. Such would be the case even if the laws of physics are written in a form in which the three constants are expressed in so called "natural" units with each constant having unit value.

70. See Smolin, L. 1997. *op. cit.*

71. Turing, A.M. 1936. *op. cit.* 135. (footnote).

72. Volchan, S.B. 2002. *op. cit.* 62.

73. Volchan, S.B. 2002. *op. cit.* 58-61.

74. Chaitin, G.J. 2001. *op. cit.* 131-134.

75. The locality constraint on the data register also ensures that the programs upon which *U* operates are self-delimiting without necessitating the addition of a special empty string symbol to delimit one program from another on the tape. As we saw in §V, all programs must be self-delimiting in order for the halting probability, Ω, to be defined. See also Volchan, S.B. 2002. *op. cit.* 58.

76. Bennett, C.H. 1982. "The Thermodynamics of Computation—a Review." *Int. J. Theor. Phys.* 21. 905-940.

77. The deconstruction of a language into prefix-free words is similar to the resolution of a vector fields into orthogonal basis vectors.

78. Volchan, S.B. 2002. *op. cit.* 51.

79. Volchan, S.B. 2002. op. cit. 52.

80. Scoville, A.E. 1995. "First Order Dark Matter and the Effectiveness of Mathematics in the Natural Sciences." in Greenberger, D.M. & Zeilinger, A. ed. 1995. *Fundamental Problems in Quantum Theory: A Conference in Honor of Professor John A. Wheeler. Vol. 755, Annals of the New York Academy of Sciences*: 896f.

81. See Freedman, W.L. 2002. "The Measure of Cosmological Parameters." arXiv:astro-ph/0202006v1:§3.

82. Reiss, A.G. *et. al.* 1998. "Observational Evidence from Supernovae for and Accelerating Universe and a Cosmological Constant." arXiv:astro-ph/9805201 v1.

83. Turner, M.S. 2001. "Dark Energy and the New Cosmology." arXiv:astro-ph/0108103.

84. Since the cold dark matter proposed here comprises only the past states of normal matter and not the matter itself, the large cosmological CDM fraction does not violate Big-Bang nucleosynthesis constraints on the baryon fraction of $u_0$. Non-neutrino leptons were not included in my computations since they form only a tiny proportion of the matter content of $u_0$.

85. The only escape here would be the possibility that there are sufficient black holes in $u_0$ tearing off portions of the trailing tape in such quantity as to equal exactly the demand for blank tape by $u_0$. Available evidence suggests that the black hole density in $u_0$ is insufficient by several orders of magnitude to constitute all of the estimated CDM in $u_0$. Even if the evidence were supportive, the same basic mechanism for the supply of blank tape from undecidable states of *E* remains as described in the body of this paper. The only difference is that $u_0$ and its undecidable states would form a closed loop. Such a closed loop violates Gödel's First Theorem.

86. Turner, M.S. 2001. *op. cit.*: 1.

87. Wigner, E.P. 1960. *op. cit.*:1.

88. There is the possibility that the proton might decay and, therefore that the hydrogen atom would disintegrate. That decay has never been observed and, in any case, the half-life of a proton is known to be >$10^{31}$ years.

89. Everett, H. 1957. *Rev. Mod. Phys.* 29, 454.

90. Linde, A. 1994. "The Self-Reproducing Inflationary Universe." *Scientific American*. November, 1994. 48-55.

91. Popper, K.R. 1965. *Conjectures and Refutations*, Routledge & Keegan Paul, Ltd. London.

92. On this point see Martin Davis's contrast between the mental materialism of John R. Searle and Roland Penrose and the mathematical platonism of Kurt Gödel. Davis, M. 2000. *The Universal Computer*. W.W. Norton & Co., Inc. New York, NY. 207.